\documentclass[aps,reprint,nofootinbib,superscriptaddress]{revtex4-1}
\pdfoutput=1

\usepackage{amsfonts}
\usepackage{amsmath}
\usepackage{hyperref}
\usepackage{amssymb}
\usepackage[english]{babel}
\usepackage{graphicx}
\usepackage{epsfig}
\usepackage{bm}
\usepackage{verbatim}
\usepackage[utf8]{inputenc}
\usepackage{color}
\usepackage{subfig}
\usepackage[normalem]{ulem}

\usepackage[capitalize]{cleveref}

\usepackage{mathtools}

\usepackage{hyperref} %links para las referencias
\hypersetup{colorlinks=true, linkcolor=red, urlcolor=red, citecolor=RoyalBlue}
\usepackage[dvipsnames]{xcolor}

\begin{document}

\title{{\color{RoyalBlue}When} $\tan \beta$ {\color{RoyalBlue} meets} {\color{red} all the mixing angles}}

\author{A. E. C\'{a}rcamo Hern\'{a}ndez}
\email{antonio.carcamo@usm.cl}
\affiliation{Universidad T\'{e}cnica Federico Santa Mar\'{\i}a,\\
Centro Cient\'{\i}fico-Tecnol\'{o}gico de Valpara\'{\i}so,\\
Casilla 110-V, Valpara\'{\i}so, Chile}

\author{C. O. Dib}
\email{claudio.dib@usm.cl}
\affiliation{Universidad T\'{e}cnica Federico Santa Mar\'{\i}a,\\
Centro Cient\'{\i}fico-Tecnol\'{o}gico de Valpara\'{\i}so,\\
Casilla 110-V, Valpara\'{\i}so, Chile}

\author{U. J. Saldana-Salazar}
\email{ulises.saldana@mpi-hd.mpg.de}
\affiliation{Max-Planck-Institut f\"ur Kernphysik,\\
Postfach 103980, D-69029 Heidelberg, Germany}

\date{\today }

\begin{abstract}
\noindent
Models with two-Higgs-doublets and natural flavour conservation contain $\tan \beta = v_2 / v_1$
as a physical parameter. We offer here a generalization of a recently proposed idea where
only the Cabibbo angle, $\theta_\text{c} \simeq 0.22$, was related to $\tan \beta$ by virtue of the 
$\mathbb{D}_{4}$  dihedral symmetry group. The original proposal consisted of a massless
first generation of quarks and no mixing with the third generation. 
In our case, through the addition of a third Higgs doublet with 
a small vacuum-expectation-value but very large masses, thus later decoupling,
all quarks  become massive and quark mixing is fully reproduced. In fact, all quark mixing angles 
are expressed in terms of $\tan \beta$ and one recovers trivial mixing 
in the limit $\beta \rightarrow 0$. We also explore the consequences in
lepton mixing by adopting a type I seesaw mechanism with three heavy 
right-handed neutrinos. 
\end{abstract}

\maketitle

\section{Introduction}
\label{sec:intro}
\noindent
Minimal scalar extensions of the standard model (SM) tackle the possibility of having
more than one fundamental scalar in Nature. However, as their construction does not
necessarily involve consideration of flavour symmetries, in general they have a large amount of arbitrariness. A part of this arbitrariness is represented 
by basis-dependent parameters which by definition are non-physical. Interestingly, when
flavour symmetries are invoked, some of these parameters survive and become physical.
Take for example the two-Higgs-doublet model (2HDM) \cite{Lee:1973iz,Branco:2011iw} in its most general scenario. Then consider
both doublets, $\Phi_1$ and $\Phi_2$, to have the same quantum numbers, thus making them identical.
By allowing the neutral components of both scalar doublets
to acquire vacuum expectation values (VEVs), 
in general the latter must fulfill the condition $ v_1^2+v_2^2 = v^2 = (174 \text{ GeV})^2$. Instead of 
$\{v_1,v_2\}$ it is equivalent to employ $\{v,\beta\}$ with $\tan \beta = v_2/v_1$. It could seem
that the angle is physical and provides a measure to distinguish between the two identical 
scalar doublets. 
However, this quantity is basis-dependent, as the kinetic terms in the scalar sector
are left invariant under global $2\times 2$ unitary transformations, and any such linear combination
is an equally valid choice; that is, there is no preferred basis.

A preferred basis is \textit{only} singled out by first imposing a certain symmetry (gauge, global, or discrete). 
The general scalar potential then reduces to a particular form. In particular, when using the reflection 
symmetry $\mathbb{Z}_2$, the non-physical parameter $\tan \beta$ can then be defined with respect to 
this basis and thereby promoted to a physical parameter. Additionally, if symmetry-breaking effects are
allowed, the identification of this parameter as physical gets more subtle. For a thorough discussion on the
physical meaning of $\tan \beta$, see Ref.~\cite{Haber:2006ue}.

In 2HDMs with $\mathbb{Z}_2$, there is natural flavour conservation~\cite{Paschos:1976ay,Glashow:1976nt}, 
that is, absence of flavour-changing-neutral-currents (FCNCs) at tree- and loop-level. Interestingly, 
the Yukawa interactions get parametrised by the corresponding Yukawa couplings and $\tan \beta$. 
One could thus naturally wonder if fermion mixing has anything to do with this parameter.
This possibility was realized only 
recently~\cite{Das:2019itj}. There it was found that by enlarging
the discrete symmetry $\mathbb{Z}_2$ to $\mathbb{D}_4 \simeq \mathbb{Z}_4 \rtimes \mathbb{Z}_2$ and 
by a judicious assignment of the quarks and the two scalars to 
irreducible representations of $\mathbb{D}_4$~\cite{Ishimori:2010au}, 
then the Cabibbo angle, 
$\theta_c \simeq 0.22$, can be directly related to $\beta$ as:
\begin{equation} \label{eq:DasRelation}
\theta_c = 2\beta \; .
\end{equation}
The proposal in \cite{Das:2019itj} is a first attempt where the first generation of quarks remains massless and there is no
allowed mixing with the third quark generation. It is our goal here to offer a complete framework where all quarks are massive and their mixings are consistent with the most 
up-to-date global fits,  and similarly for the leptons. \\

{There has already been various efforts towards the construction of a successful $\mathbb{D}_4$ flavor model~\cite{Grimus:2003kq,Grimus:2004rj,Blum:2007jz,Adulpravitchai:2008yp,Hagedorn:2010mq,Meloni:2011cc}. For example, from a rather general point of view Ref.~\cite{Blum:2007jz} offers a thorough discussion of the major implications of using dihedral groups. Unfortunately, none of them explored the possibility of relating the masses and mixing parameters to $\tan \beta$. Our purpose here is to explore this relation as a direct consequence of $\mathbb{D}_4$. Therefore, we will not consider further auxiliary symmetries that could reduce the number of free parameters. \\}

This letter is organized as follows. In Sec.~\ref{sec:model} we describe the model. 
In Sec.~\ref{sec:scalarsector} we discuss the scalar potential and show how
to produce hierarchical VEVs. In Secs.~\ref{sec:quarksector} and~\ref{sec:leptonsector} 
we show how the mixing angles can be related to $\tan \beta$. { We discuss the main features and phenomenological consequences of the model in Secs.~\ref{sec:pheno}, \ref{sec:photons}, \ref{sec:HeavyScalar}, and \ref{sec:discussion}. Finally, our conclusions are stated in Sec.~\ref{sec:conclusions}. To keep the discussion short we have delegated all technical details to Appendices.}

\section{Model Description}
\label{sec:model}
\noindent
$\mathbb{D}_4 \simeq \mathbb{Z}_4 \rtimes \mathbb{Z}_2$ is the 
symmetry group of a square. It is a discrete, non-abelian group. Two independent symmetry transformations characterize it: reflections and $\pi/2$ rotations. It has four singlet representations and one doublet, here denoted as 
$\mathbf{1}_{++}$, $\mathbf{1}_{+-}$, $\mathbf{1}_{-+}$, $\mathbf{1}_{--}$ and $\mathbf{2}$,
respectively. The multiplication rules are shown in Appendix~\ref{app:rules}.
\\

In contrast to Ref.~\cite{Das:2019itj}, we { use a different basis~\cite{Ishimori:2010au,Meloni:2011cc} for the generators of the two-dimensional representations\footnote{ This choice directly impacts the tensorial products,
as shown in Eq.~\eqref{eq:TensorialProductsD4}.}
\begin{equation}
	{\bf a} = \begin{pmatrix}
	i & 0 \\
	0 & -i
	\end{pmatrix} \qquad \text{and} \qquad
	{\bf b} = \begin{pmatrix}
	0 & 1 \\
	1 & 0
	\end{pmatrix} \;,
\end{equation}
where ${\bf a}$ and ${\bf b}$ denote the generators of $\pi/2$ rotations and reflections,
and are order four and two, respectively: ${\bf a}^4 = {\bf 1}_{2\times 2} = {\bf b}^2$. {Notice that our generators are complex. Since the representations are real, there is a unitary matrix ${\bf \sigma}_1 $ that connects the generators to the complex conjugates:  
\begin{equation}
{\bf \sigma}_1 = \begin{pmatrix}
0 & 1 \\
1 & 0
\end{pmatrix} \;.
\end{equation}
As a consequence, for any given flavor doublet, ${\bf g} = (g_1,g_2)^T$, when considering the conjugate, it is the combination $\sigma_1 {\bf g}^*$ that transforms as a doublet, not simply ${\bf g}^*$. {Furthermore, realize the main advantage of choosing the aforementioned complex generators: the  up and down components of the flavor doublet have definite $\mathbb{Z}_4$ charge which could later turn to be useful if breaking the flavor symmetry into a particular subgroup. For further details see Ref.~\cite{Ishimori:2010au}.}} \\

We make the following representation assignments in the quark sector (we use the subindex ``$D$'' to denote the $\mathbb{D}_4$ doublets):
\begin{equation}
\begin{gathered}
Q_{3L} \sim  \,\mathbf{1}_{++} \;, \qquad u_{3R} \sim  \,\mathbf{1}_{++} \;, \qquad d_{3R} \sim \,\mathbf{1}_{ -+} \;, \\
 Q_{DL} =  \begin{pmatrix}Q_{1L}\\Q_{2L}\end{pmatrix} \sim 
\mathbf{2}\,, \quad
 u_{DR} = \begin{pmatrix}u_{1R}\\u_{2R}\end{pmatrix} \sim 
\mathbf{2}\,,  \\
d_{DR} = \begin{pmatrix}d_{1R}\\d_{2R}\end{pmatrix}  \sim 
\mathbf{2} \;,
\end{gathered}
\end{equation}
whereas in the lepton sector we assign:
\begin{equation}
\begin{gathered}
\ell_{3L} \sim  \,\mathbf{1}_{++} \;, \qquad N_{3R} \sim  \,\mathbf{1}_{++} \;, \qquad e_{3R} \sim \,\mathbf{1}_{ -+} \;, \\
 \ell_{DL} =  \begin{pmatrix}\ell_{1L}\\ \ell_{2L}\end{pmatrix} \sim 
\mathbf{2}\,, \quad
 e_{DR} = \begin{pmatrix}e_{1R}\\e_{2R}\end{pmatrix} \sim 
\mathbf{2}\,,  \\
N_{DR} = \begin{pmatrix}N_{1R}\\N_{2R}\end{pmatrix}  \sim 
\mathbf{2} \;.
\end{gathered}
\end{equation}
Here we use the notation of Ref.~\cite{Ishimori:2010au} for the one-dimensional representations, namely $1_{b,ab}$ ($i.e.$ the signs indicate the transformation of the field under  $b$ and $ab$, respectively).

Our choice allows a later reinterpretation of the model as having the appearance of
a 2HDM with softly-broken natural flavour conservation. Notice that we are also considering $\mathbb{D}_4$-assignments of the leptonic fields, which are not included in the work of Ref.~\cite{Das:2019itj}, as is only focused on the description of the quark sector.

On the other hand, the scalar sector, which is composed of three Higgs doublets, has the assignments
\begin{align}
\begin{split}
\Phi_{D} = \begin{pmatrix}\Phi_{1}\\ \Phi_{2}\end{pmatrix}  \sim 
\mathbf{2} 
 \qquad \text{and} \qquad
\Phi_S \sim \mathbf{1}_{  -+}  \;.
\end{split}%
\end{align}
There are other possibilities $\Phi_S \sim \{{\bf 1}_{++}, {\bf 1}_{--},{\bf 1}_{+-} \}$ (for more details see Appendix~\ref{app:scenarios}), however, not all of
the resulting mass matrices give the correct masses and mixings. 
Note that the scalar doublet $\Phi_S$ is a necessary element as by virtue of it we introduce mixing with the third generation and a non-zero mass for the first fermion family. Moreover, as seen later, it can also be used to give mass to the bottom quark and tau lepton, opening the parameter space to more viable solutions. In what follows, we also use $\widetilde\Phi = i\tau_2 \Phi^\ast$, where $\tau_2$ is the Pauli matrix in $SU(2)_L$ space. 

The Yukawa Lagrangian for the quark sector is ${\cal L}_Y^{Q} =  {\cal L}_Y^{(u)}+{\cal L}_Y^{(d)}$,
where {
\begin{align}
\begin{split}
-{\cal L}_Y^{(u)} = & \;  y_1^u (\overline{Q}_{2L}  u_{1R}  - \overline{Q}_{1L}  u_{2R} ) \widetilde{\Phi}_S  \\ & + y_2^u (\overline{Q}_{1L} \widetilde{\Phi}_2 + \overline{Q}_{2L} \widetilde{\Phi}_1) u_{3R} \\
& + y_3^u \overline{Q}_{3L}(\widetilde{\Phi}_1 u_{1R} + \widetilde{\Phi}_2 u_{2R}) + \text{ H.c.} \;, 
\\
-{\cal L}_Y^{(d)} = & \; y_1^d (\overline{Q}_{2L}  d_{1R} - \overline{Q}_{1L}  d_{2R}) {\Phi}_S \\
& +  y_2^d (\overline{Q}_{2L} {\Phi}_1 - \overline{Q}_{1L} {\Phi}_2) d_{3R} \\ 
& + y_3^d \overline{Q}_{3L}({\Phi}_1 d_{2R} + {\Phi}_2 d_{1R})  \\
& + y^d_4 \overline{Q}_{3L} {\Phi}_S d_{3R} + \text{ H.c.} \; ,
\label{eq:LagQuark}
\end{split}%
\end{align}   }
whereas for the lepton sector is ${\cal L}_Y^{\ell} =  {\cal L}_Y^{(e)}+{\cal L}_Y^{(\nu)}$, with {
\begin{align}
\begin{split}
-{\cal L}_Y^{(e)} = & \; y_1^e (\overline{\ell}_{2L}  e_{1R} - \overline{\ell}_{1L}  e_{2R}) {\Phi}_S \\
& + y_2^e (\overline{\ell}_{2L} {\Phi}_1 - \overline{\ell}_{1L} {\Phi}_2) e_{3R} \\ 
& + y_3^e \overline{\ell}_{3L}({\Phi}_1 e_{2R} + {\Phi}_2 e_{1R} ) \\
& + y_4^e \overline{\ell}_{3L} \Phi_S e_{3R} +
\text{ H.c.} \;,
\\
-{\cal L}_Y^{(\nu)} = & \; y_1^\nu ( \overline{\ell}_{2L}  N_{1R} - \overline{\ell}_{1L}  N_{2R}) \widetilde{\Phi}_S  \\
& + y_2^\nu (\overline{\ell}_{1L} \widetilde{\Phi}_2 + \overline{\ell}_{2L} \widetilde{\Phi}_1) N_{3R}  \\
& + 
y_3^\nu \overline{\ell}_{3L}(\widetilde{\Phi}_2 N_{2R}+ \widetilde{\Phi}_1 N_{1R}) \\
& + \frac{M_2}{2} \overline{N_{3R}^c} N_{3R} 
\\
& + \frac{M_1}{2} (\overline{N_{1R}^c} N_{2R} + \overline{N_{2R}^c} N_{1R}) + \text{ H.c.} \; .
\label{eq:LagLep}
\end{split}%
\end{align}      
Note that we have three complex parameters for the up quark mass matrix, and also three for the Dirac neutrino mass matrix, while we have four complex parameters for the down quark and the charged lepton mass matrices; and finally, two real parameters for the mass matrix of the Majorana neutrinos, $N_{1R}$ and $N_{2R}$. Now,} a phase field redefinition can make all the phases in the up quark mass matrix to be zero while only two and three in the down quarks and charged leptons, respectively, as obtained in Appendix~\ref{app:complexphases}. Therefore, we choose to keep as complex parameters, in the down quark and charged lepton mass matrix, only $\{y_3^d, y_4^d\}$ and $y_4^e$.  Thus we are left with 7 real magnitudes and 2 relevant complex phases in the quark sector, while 9 real magnitudes and 4 complex phases in the lepton sector (including the two Majorana masses). The 9 (13) arbitrary parameters in the quark (lepton) sector must describe 10 (12) mass and mixing parameters. Hence, it is only in the quark sector that we expect the strongest correlations.

\section{The scalar potential}
\label{sec:scalarsector}
\noindent
The most general, renormalizable, and $\mathbb{D}_4$-symmetric scalar potential is given by  
\begin{align} 
\begin{split}
V = & \, \mu_D^2 ({ \Phi_1^\dagger \Phi_1 + \Phi_2^\dagger \Phi_2}) 
+ \mu_S^2 \Phi_S^\dagger \Phi_S  
+ \frac{\lambda_{1}}{2} (\Phi_S^\dagger \Phi_S)^2 
\\ &
+ \frac{\lambda_{2}}{2} (\Phi_1^\dagger \Phi_2 + \Phi_2^\dagger \Phi_1)^2 
+ \frac{\lambda_{3}}{2} (\Phi_1^\dagger \Phi_2 - \Phi_2^\dagger \Phi_1)^2  
\\ &
+ \frac{\lambda_{4}}{2} (\Phi_1^\dagger \Phi_1 + \Phi_2^\dagger \Phi_2)^2 
+ \frac{\lambda_{5}}{2} (\Phi_1^\dagger \Phi_1 - \Phi_2^\dagger \Phi_2)^2
\\ &
+ \frac{\lambda_{6}}{2}[ { (\Phi_2^\dagger \Phi_S)(\Phi_S^\dagger \Phi_2) + (\Phi_1^\dagger \Phi_S)(\Phi_S^\dagger \Phi_1) }]
 \\ & 
 + \frac{1}{2}[ { -}\lambda_{7}(\Phi_2^\dagger \Phi_S)(\Phi_1^\dagger \Phi_S) { -} \lambda_{7}(\Phi_1^\dagger \Phi_S)(\Phi_2^\dagger \Phi_S) + \text{ H.c.} ]
 \\ &
+ \frac{\lambda_{8}}{2}({ \Phi_1^\dagger \Phi_1 + \Phi_2^\dagger \Phi_2)(\Phi_S^\dagger \Phi_S }) \;.
\end{split}
\end{align} 
Because of Hermiticity, all parameters except for $\lambda_7$ are real. 
However, a phase redefinition of the fields can absorb the phase of $\lambda_7$ { together with the global minus sign}. Therefore, the potential is also invariant under the Charge-Parity (CP) discrete transformation.

We assume $\mu_D^2 < 0$ while $\mu_S^2 >0$, such that
\begin{align}
\langle \Phi_D^0 \rangle = \begin{pmatrix}
	v_1 \\
	v_2
\end{pmatrix} \qquad \text{and} \qquad 
\langle \Phi_S^0 \rangle = 0 \;.
\end{align}
As long as the flavour symmetry is not broken, the minimum conditions: 
\begin{equation}
\begin{gathered}
v_1 \left[ \mu^2_D+\left(2\lambda_2+\lambda_4-\lambda_5\right)v_2^2+\left(\lambda_4+\lambda_5\right){v^2_1} \right] =0 \\
v_2 \left[ \mu^2_D+\left(2\lambda_2+\lambda_4-\lambda_5\right)v_1^2+\left(\lambda_4+\lambda_5\right){v^2_2} \right] =0
\end{gathered}
\end{equation} 
enforce { two independent solutions: the symmetric
limit, $v_1 = v_2$, or one in which one of the two VEVs is zero while the other is equal to $\sqrt{-\mu_D^2/(\lambda_4 + \lambda_5)}$. 

In the following, we choose the latter possibility. Now,} if we wish to explore the case where { the null} VEV in the flavour doublet is { no longer zero but still} rather small, i.e. $v_2 \sim {\cal O}(10^{-2}-10^{-1}) \, v_1$, we need to softly-break the symmetry by introducing
\begin{align} {
-\mu_{12}^2 (\Phi_1^\dagger \Phi_2 + \Phi_2^\dagger \Phi_1) \;,  }
\end{align}
where {to conserve CP} we assume $\mu_{12}^2$ to be real { and we require $\mu_{12}^2 \ll \mu_D^2$ and also $\mu_{12}^2 > 0$. Furthermore, it is also necessary to break the mass scale between $\Phi_1$ and $\Phi_2$. For this purpose, we add $\mu_2^2 \Phi_2^\dagger \Phi_2$.
} 

{ Realize that} as we are also interested in \textit{inducing} a VEV
in the flavour singlet isodoublet, $\Phi_S$, we need to add an extra
soft-breaking term. Now, to make its VEV smaller than $v_2$ we choose $\Phi_2$ and
not $\Phi_1$ as the one responsible for inducing it. The complete $non-symmetrical$ expression therefore reads
\begin{align}
\begin{split}
V_\text{non-symm} = & \, { \mu_2^2 \Phi_2^\dagger \Phi_2
 -\mu_{12}^2 (\Phi_1^\dagger \Phi_2 + \text{ H.c.})} \\ 
& - \frac{\mu_{S2}^2}{ 2} (\Phi_S^\dagger \Phi_2 + \text{ H.c.})\;.
\end{split}
\end{align}
This choice allows us to write the following relations 
\begin{equation} \label{eq:VEVs}
\begin{gathered}
 v_1 = \sqrt{\frac{-\mu_D^2}{\lambda_4 + \lambda_5}} \;, \qquad
 v_2 \simeq \frac{\mu_{12}^2 v_1}{\mu_2^2 + \mu_D^2 + (2\lambda_2 + \lambda_4 - \lambda_5) v_1^2} \;, \\
v_s \simeq \frac{\mu_{S2}^2 v_2 }{2\mu_S^2+(\lambda_6+\lambda_8)v_1^2} \;.
\end{gathered}
\end{equation} 
If we consider 
$\lambda_k \sim {\cal O}(1)$, and
 $|\mu_D| \sim {\cal O}(100) \text{ GeV}$, $\mu_{2} \sim {\cal O}(100) \text{ GeV}$, $\mu_{12} \sim {\cal O}(10) \text{ GeV}$, 
$\mu_{S2} \sim {\cal O}(0.1 - 1)  \text{ TeV} $, and $\mu_S \sim {\cal O}(1) \text{ TeV}$ then we expect {
$v_1 \sim {\cal O}(100) \text{ GeV}$, $v_2 \sim {\cal O}(10) \text{ GeV}$, and $v_s \sim {\cal O}(1) \text{ GeV}$}.
In other words, $v_2 = \epsilon v_1$ and $v_s = \epsilon^2 v_1$ with $\epsilon \sim 10^{-1}$.
Now, note that this hierarchy in the VEVs allows us to say that to a very good degree of approximation
\begin{align}
v^2 = v_1^2 + \epsilon^2 v_1^2 + \epsilon^4 v_1^2 \approx v_{1}^2 + v_{2}^2 + {\cal O}(\epsilon^4) \;,
\end{align}
and we are still able to write
\begin{align}
\tan \beta \simeq \frac{v_2}{v_1} \;.
\end{align}
With this definition we can reexpress the small VEVs in the following way
\begin{equation}
\begin{gathered}
v_2 \simeq v \tan \beta \;, \\
v_s \simeq \frac{\mu_{S2}^2  v \tan \beta }{{ 2\mu_s^2+v^2(\lambda_6+\lambda_8) }} \;,
\end{gathered}
\end{equation}
where we have already considered $v_1 \simeq v$ and {
\begin{equation} 
\tan \beta \simeq \frac{\mu_{12}^2 }{\mu_2^2 + \mu_D^2+(2\lambda_2+\lambda_4-\lambda_5)v^2} \; .
\end{equation} }

To illustrate the order of the scalar masses we consider a simple benchmark scenario where
$\lambda_{2,3,6,7,8} = 0$, $\lambda_4 = \lambda_5$, and $\lambda_{1,4,5} \sim {\cal O}(1)$. This scenario has the great advantage of simplifying Eq.~\eqref{eq:VEVs} to
\begin{equation} {
\begin{gathered}
v_1 = \sqrt{\frac{{-}\mu^2_D}{2\lambda_5}} \;, \qquad
\frac{v_2}{v_1} = \frac{ \mu_{12}^2}{\mu_2^2+\mu_D^2} \;, \qquad
\frac{v_s}{v_2} = \frac{\mu_{S2}^2}{2\mu_S^2} \;,
\end{gathered} }
\end{equation}
and allowing a qualitative analysis of both the scalar masses and the hierarchical VEVs
in terms of the same set of parameters { $\{\mu_D^2,\mu_2^2,\mu_{12}^2,\mu_S^2,\mu_{S2}^2,\lambda_1,\lambda_5\}$}.
{ The complete and exact mass matrices are shown in Appendix~\ref{app:scalar}.}
The scalar mass matrices, { for this case}, then read
\begin{subequations}
\begin{eqnarray}
{\bf m}^2_{CP-\text{even}}  &  \simeq &  \begin{pmatrix}
-2\mu_D^2 &- \mu_{12}^2 & 0\\
-\mu_{12}^2  &   \mu_2^2 + \mu_D^2    & -\mu_{S2}^2 /2\\
0 & -\mu_{S2}^2 /2 & \mu_S^2
\end{pmatrix}  ,\quad
\\
{\bf m}^2_{CP-\text{odd}}  & \simeq  & \begin{pmatrix}
0 & -\mu_{12}^2 & 0\\
-\mu_{12}^2  & \mu_2^2  + \mu_D^2 & -\mu_{S2}^2 /2 \\
0 & -\mu_{S2}^2 /2 & \mu_S^2
\end{pmatrix} ,
\\
{\bf m}^2_{\text{charged}}  & \simeq  & \begin{pmatrix}
0 & -\mu_{12}^2 & 0\\
-\mu_{12}^2  & \mu_2^2  + \mu_D^2 & -\mu_{S2}^2 /2 \\
0 & -\mu_{S2}^2 /2 & \mu_S^2
\end{pmatrix} , 
\end{eqnarray}         
\end{subequations}      
where we have neglected small contributions when appropriate.
The CP-odd and charged scalar mass matrices are consistent with only one massless state.
{ Furthermore, in this benchmark scenario,}
 {the physical CP odd neutral and electrically charged scalars are exactly degenerate, which is an scenario favoured by electroweak precision tests \cite{Hernandez:2015rfa}.} It is then straightforward to obtain that the lightest scalar state of mass $125 \text{ GeV}$ %
 while all the heavy states (neutral and charged) have masses %o
 in agreement with all the current experimental constraints. {In fact, we have performed a numerical analysis of the scalar sector obtaining a large number of solutions for the scalar masses consistent with experimental bounds. In our analysis we have required that the mass for the lightest CP even neutral scalar state to be in the $3\sigma$ experimentally allowed range $124.96$ GeV$\leqslant m_h \leqslant$ $125.8$ GeV \cite{Sirunyan:2020xwk}, whereas for the heavy CP even neutral scalar masses we require them to be larger than $200$ GeV, as done in \cite{Hernandez-Sanchez:2020vax}. The masses for the physical CP odd and electrically charged scalars are required to be larger than their lower experimental bounds of $93.4$ GeV and $90$ GeV, respectively \cite{Tanabashi:2018oca}. 
We have found that the heavy scalar masses feature linear correlations as shown in Figure \ref{scalarcorrelations}.}%,

\begin{figure}[h]
	\resizebox{7.5cm}{4cm}{\vspace{-2cm}%
		\includegraphics{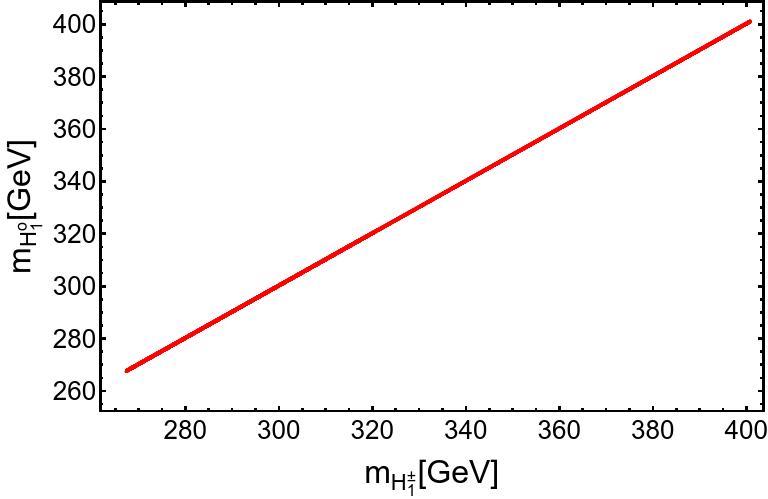}		
	}\vspace{0cm}
	\resizebox{7.5cm}{4cm}{\vspace{-2cm}%
		\includegraphics{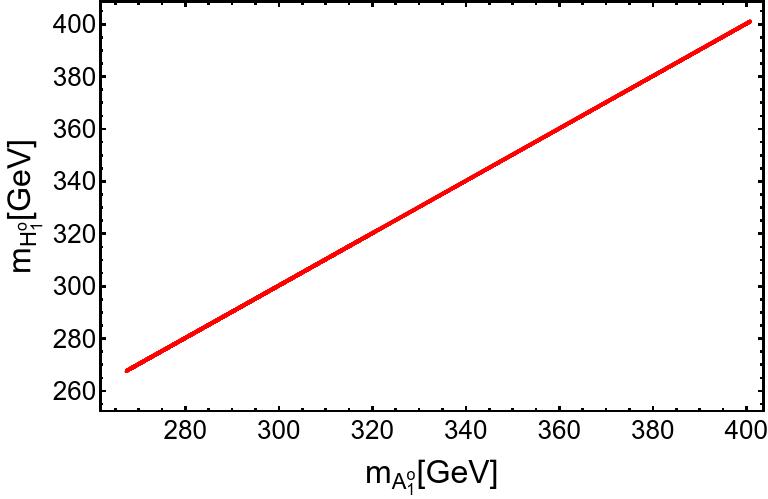}		
	}\vspace{0cm}
	\resizebox{7.5cm}{4cm}{\vspace{-2cm}%
		\includegraphics{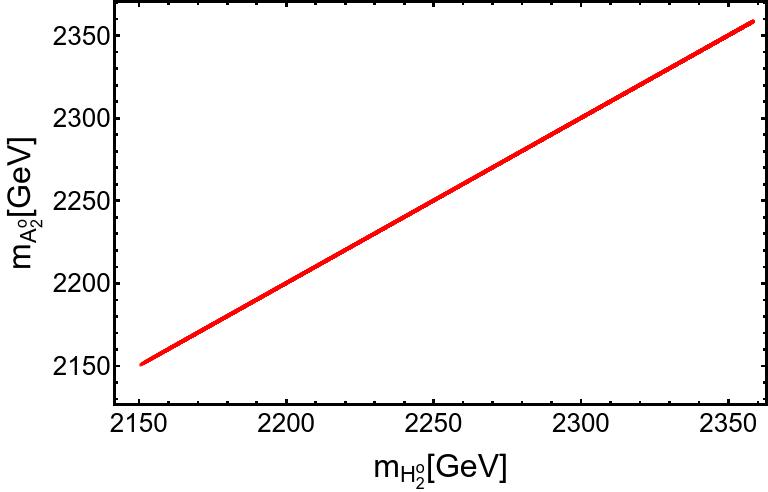}		
	}\vspace{0cm}
\resizebox{7.5cm}{4cm}{\vspace{-2cm}%
	\includegraphics{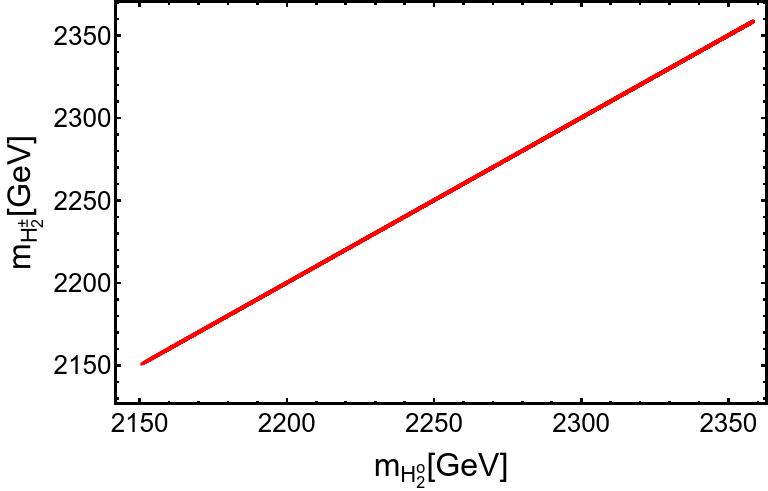}}
	\caption{Correlations between the heavy scalar masses.}
	\label{scalarcorrelations}
\end{figure}

\section{Quark masses and mixings}
\label{sec:quarksector}
\noindent
The mass matrices in the quark sector are
\begin{align} \label{eq:Mu}
{\bf M}_u & =
\begin{pmatrix}
0 & -y_1^u v_s & y_2^u v_2 \\
y_1^u v_s & 0 & y_2^u v_1 \\
 y_3^u v_1 & y_3^u v_2 &0 
\end{pmatrix} \;,
\\ \label{eq:Md}
{\bf M}_d & =
\begin{pmatrix}
0 & -y_1^d v_s & -y_2^d v_2 \\
y_1^d v_s & 0 & y_2^d v_1 \\
y_3^d v_2 e^{i \delta_3} & y_3^d v_1 e^{i \delta_3} & y_4^d v_s e^{i \delta_4}
\end{pmatrix} \;,         
\end{align}    
where in this notation all Yukawa couplings of the form $y_k^q$ are real and positive.
In Appendix~\ref{app:scenarios}, we have fully considered all the different possible $\mathbb{D}_4$ assignment scenarios of the kind ${\bf 2} + {\bf 1}_x$ with three families of fermions and Higgs doublets.

We calculate the unitary transformations diagonalizing the mass matrices from the 
hermitian mass matrix product, { ${\bf H} = {\bf M} {\bf M}^\dagger$},
\begin{footnotesize}
\begin{equation} \label{eq:quarkmassmatrix} 
\begin{gathered}
{\bf H}_u  =
\begin{pmatrix}
(y_1^u)^2 v_s^2 + (y_2^u)^2 v_2^2 & (y_2^u)^2 v_1 v_2 & -y_1^u y_3^u v_2 v_s \\
(y_2^u)^2 v_1 v_2 & (y_1^u)^2 v_s^2 + (y_2^u)^2 v_1^2 & y_1^u y_3^u v_1 v_s \\
 -y_1^u y_3^u v_2 v_s & y_1^u y_3^u v_1 v_s  & (y_3^u)^2(v_1^2 + v_2^2)
\end{pmatrix} 
\;, \\
{\bf H}_d  =
\begin{pmatrix}
(y_1^d)^2 v_s^2 + (y_2^d)^2 v_2^2 & -(y_2^d)^2 v_1 v_2 & -y_1^d y_3^d v_1 v_s e^{-i \delta_3}\\
-(y_2^d)^2 v_1 v_2 & (y_1^d)^2 v_s^2 + (y_2^d)^2 v_1^2 & y_1^d y_3^d v_2 v_s e^{-i \delta_3}\\
 -y_1^d y_3^d v_1 v_s e^{i \delta_3} & y_1^d y_3^d v_2 v_s e^{i \delta_3} & (y_3^d)^2(v_1^2 + v_2^2)
\end{pmatrix} \\
+  
\begin{pmatrix}
0 & 0 & -y_2^d y_4^d v_2 v_s e^{-i \delta_4}\\
0 & 0 & y_2^d y_4^d v_1 v_s e^{-i \delta_4}\\
-y_2^d y_4^d v_2 v_s e^{i \delta_4} & y_2^d y_4^d v_1 v_s e^{i \delta_4} & (y_4^d)^2 v_s^2
\end{pmatrix} \;,
\end{gathered}   
\end{equation}    
\end{footnotesize} 
If we were to try to fit the quark masses and mixing with those contributions that remain when setting $y_4^d$ to zero, we would find out that it is not possible. In fact, in all those settings appearing in Appendix~\ref{app:scenarios} with $y_4^d = 0$ there is no viable phenomenology. The natural advantage of considering its contributions is that thanks to the amount of suppression in $v_s \sim \beta^2 v_1$, we can make them responsible for introducing the mass of the bottom quark with $y_4^d \sim {\cal O}(1)$ together with the right mixing, as shown in Appendix~\ref{app:Relating}. This observation is what allows the present framework to be viable.

Notice that when $v_s \rightarrow 0$, the up- and down-type quark
mass matrices shown in Eq.~\eqref{eq:quarkmassmatrix} are diagonalised
by the orthogonal transformation
\begin{align} \label{eq:rotation}
{\bf R}_{u(d)} = \begin{pmatrix}
\cos \beta & {\mp} \sin \beta & 0\\
{\pm} \sin \beta & \cos \beta & 0 \\
0 & 0 & 1
\end{pmatrix} \; .
\end{align}
Hence, the leading contribution to quark mixing is obtained from
\begin{align}
{\bf V}_q = {\bf R}_{u} {\bf R}_{d}^T = \begin{pmatrix}
\cos 2 \beta & -\sin 2\beta & 0\\
\sin 2\beta & \cos 2\beta & 0 \\
0 & 0 & 1
\end{pmatrix} \;,
\end{align}
from which one easily sees that Eq.~\eqref{eq:DasRelation} is satisfied up to a sign
\begin{equation} \label{eq:correctDasrelation}
\theta_c = - 2 \beta \;.
\end{equation}
In fact, we would have obtained a positive sign if the right-handed fermions of
the third family had initially exchanged their assignments, i.e. 
\begin{equation}
	u_{3R} \sim {\bf 1}_{ -+} \quad \text{and} \quad
	d_{3R} \sim {\bf 1}_{++} \;.
\end{equation}
Nevertheless, within the original relation ($\theta_c = + 2\beta$), it is not possible to find a good agreement of the mixing parameters with their experimental value. Therefore, we conclude from here that in a complete $\mathbb{D}_4$-framework with three Higgs doublets and where all quark mixing phenomena is explained, the correct relation between the Cabibbo and $\beta$ angle requires a minus sign, as in Eq.~\eqref{eq:correctDasrelation}. 
In Appendix~\ref{app:Relating}, we give all the necessary details to understand the aforementioned observation and also realize how the mixing angles get expressed in terms of $\beta$.

\begin{figure}[h]
	\resizebox{7.5cm}{4cm}{\vspace{-2cm}%
	\includegraphics{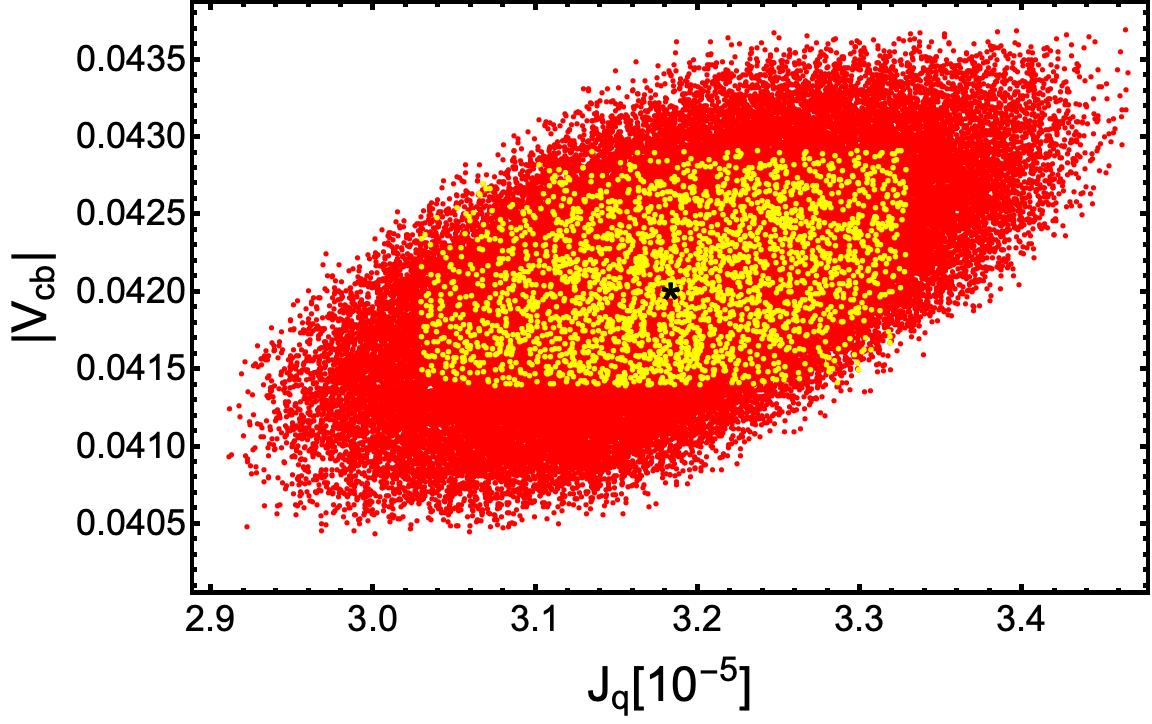}		
	}\vspace{0cm}
\resizebox{7.5cm}{4cm}{\vspace{-2cm}%
	\includegraphics{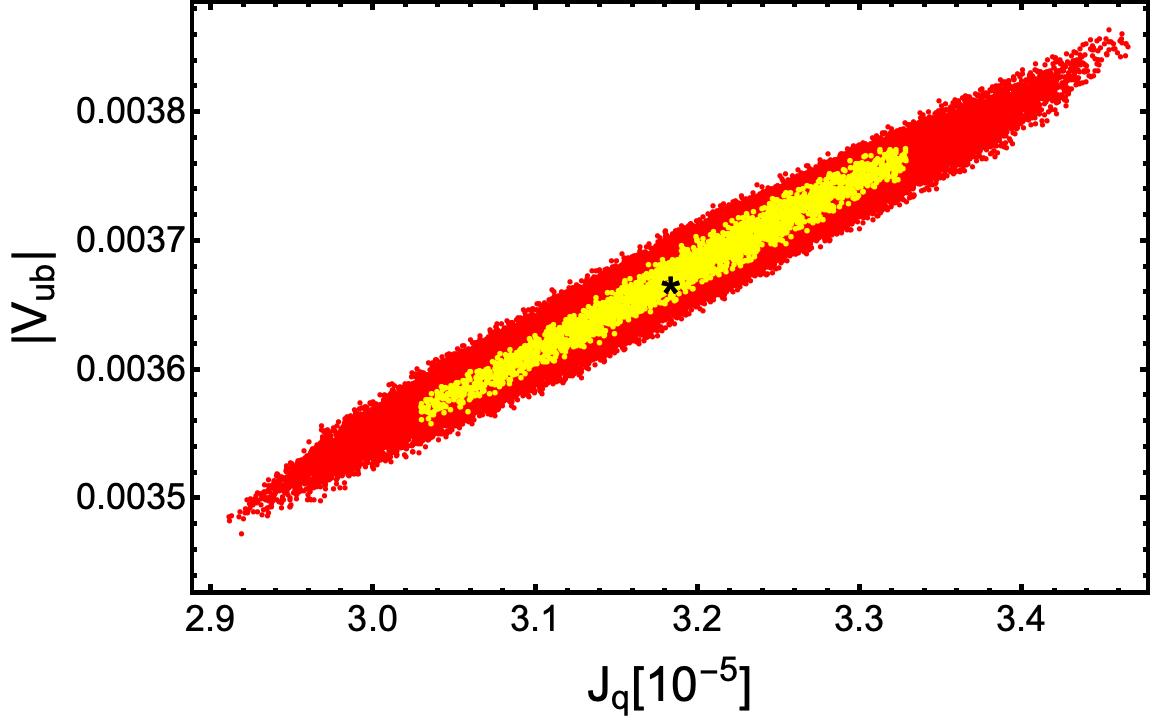}		
}\vspace{0cm}
\resizebox{7.5cm}{4cm}{\vspace{-2cm}%
	\includegraphics{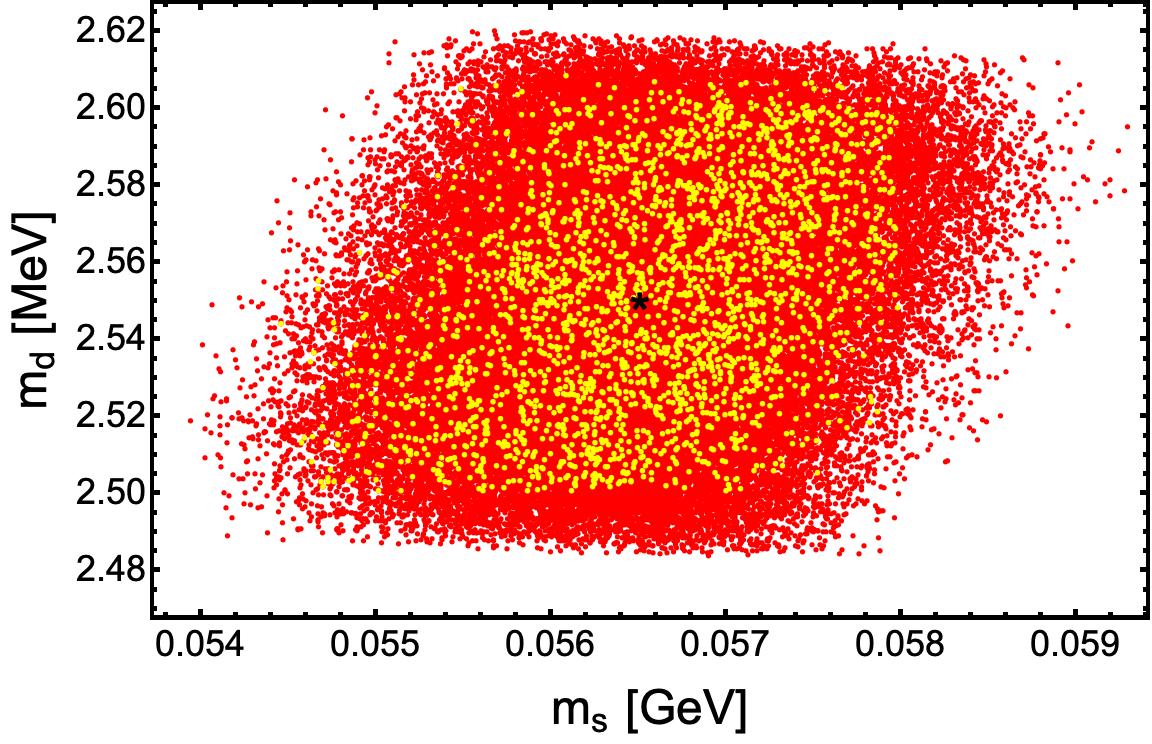}		
}\vspace{0cm}
	\caption{ Correlations of the { smallest} quark mixing angles with the Jarlskog invariant. { Additionally, we also show the correlation between the down and strange quark masses.} The red and yellow points show the 3$\sigma$ and 1$\sigma$ regions around the best-fit point given by the black star.}
	\label{Quarkcorrelations}
\end{figure}

{We have performed a numerical analysis of the quark mass matrices and found a wide region of parameter space where the obtained values of the quark mixing angles, Jarlskog invariant and the up and down type quark masses are consistent with the experimental data, see Appendix~\ref{app:chiQuarks}.  

The following best-fit values, 
\begin{equation} \label{eq:BFP} {
\begin{gathered}
    y^u_1 = 4.6764 \times 10^{-4}  , \quad
    y^u_2 = 3.5950 \times 10^{-3}  , \quad
    y^u_3 = 0.9891  , \\
    y^d_1 =  4.2428 \times 10^{-3}  , \quad
    y^d_2 = 7.59323 \times 10^{-4} , \\
    y^d_3 = 6.7225 \times 10^{-3} e^{i 3.3436} , \quad
    y^d_4 = 0.99096 e^{i 1.3945}
\end{gathered} }
\end{equation}
reproduce all the quark masses and the observed quark mixing at the $1 \sigma$ level with a quality of fit of { $\chi^2_\text{d.o.f.} = 0.10$}. The correlations of the quark mixing angles with the Jarlskog invariant are shown in Figure~\ref{Quarkcorrelations}. Interestingly, while doing the fit we find that the relation $|\theta_c| = 2|\beta|$ receives { significant contributions}. Therefore, instead of using the value $|\beta| \simeq |\theta_c| /2$ we explicitly employed { $|\beta| = 0.047$}. Further details have been delegated to Appendix~\ref{app:chiQuarks}.
}

\section{Lepton masses and mixings}
\label{sec:leptonsector}
\noindent
The Dirac mass matrices in the lepton sector are 
\begin{align} \label{eq:leptons1}
{\bf M}_\nu & =
\begin{pmatrix}
0 & -y_1^\nu v_s e^{i \delta_1^\nu}  & y_2^\nu v_2 e^{i \delta_2^\nu} \\
y_1^\nu v_s e^{i \delta_1^\nu} & 0 & y_2^\nu v_1 e^{i \delta_2^\nu} \\
 y_3^\nu v_1 e^{i \delta_3^\nu} & y_3^\nu v_2 e^{i \delta_3^\nu}&0 
\end{pmatrix} \;,
\\ \label{eq:leptons2}
{\bf M}_e & =
\begin{pmatrix}
0 & -y_1^e v_s & -y_2^e v_2 \\
y_1^e v_s & 0 & y_2^e v_1 \\
y_3^e v_2 & y_3^e v_1 & e^{i \delta_4^e }y^e_4 v_s
\end{pmatrix} \;,
\end{align}            
while the corresponding one for the right-handed neutrinos is
\begin{align}
{\bf M}_N & =
\begin{pmatrix}
0 & M_{1} & 0 \\
M_{1} & 0 & 0 \\
0 & 0 & M_2 
\end{pmatrix} \;.
\end{align}

\begin{figure}[t]
	\resizebox{7.5cm}{4cm}{\vspace{-2cm}%
		\includegraphics{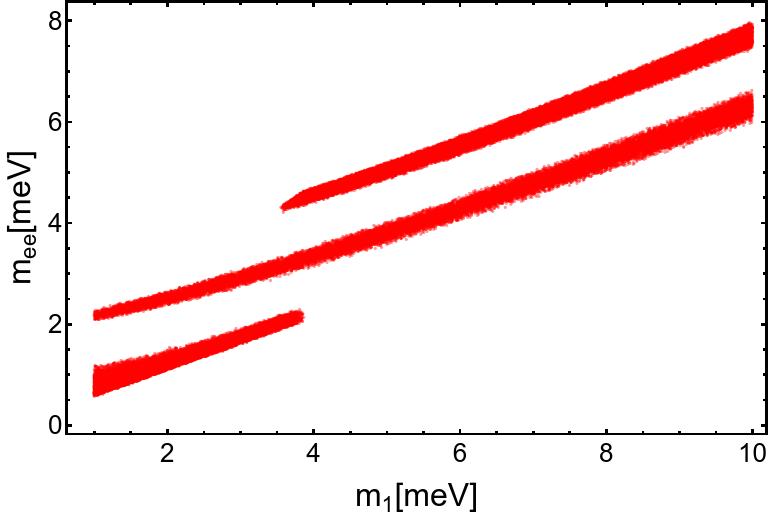}		
	}\vspace{0cm}
	\resizebox{7.5cm}{4cm}{\vspace{-2cm}%
		\includegraphics{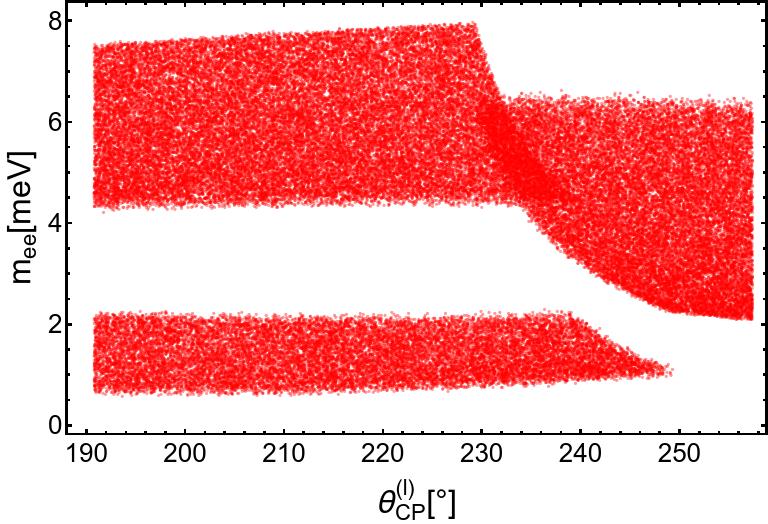}		
	}
	\caption{Correlations of the effective mass parameter of neutrinoless double beta decay with the lightest neutrino mass and leptonic CP phase for the scenario of normal neutrino mass hierarchy.}
	\label{Leptoncorrelations}
\end{figure}

Assuming that the right-handed Majorana neutrinos have masses much larger than the electroweak symmetry breaking scale, { $\{M_1,M_2\} \gg v_1$}, one has that the light active neutrino masses are generated from a type I seesaw mechanism that gives rise to an effective Majorana mass matrix
\begin{equation}
{\bf M}_{\text{maj}} \simeq - {\bf M}_\nu {\bf M}^{-1}_N {\bf M}_\nu^T \;,
\end{equation} 
The lepton mass matrices show some interesting features. In the limit $\beta \rightarrow 0$, the solar, reactor, and atmospheric mixing angles go to zero as well as the first family of lepton masses. This can be obtained from noting that, in the limit $\beta \rightarrow 0$, Eqs.~\eqref{eq:leptons1} and~\eqref{eq:leptons2} take the form
\begin{align}
{\bf M}_{e} = \begin{pmatrix}
0 & 0 & 0 \\
0 & 0 & \times \\
0 & \times & 0
\end{pmatrix},\hspace{1cm}{\bf M}_{\nu} = \begin{pmatrix}
0 & 0 & 0 \\
0 & 0 & \times \\
\times & 0 & 0
\end{pmatrix} \;,
\end{align}
thus implying that the effective Majorana mass matrix for neutrinos turns into 
\begin{equation}
{\bf M}_{\text{maj}} = \begin{pmatrix}
0 & 0 & 0 \\
0 & \times & 0 \\
0 & 0 & 0
\end{pmatrix} \;,
\end{equation}
favouring the inverted ordering case. However, in general, for $\beta \neq 0$ the hierarchical structure among the mass matrix elements strongly depends upong the heavy Majorana masses. This means that depending on $M_1\gg M_2$ or $M_1 \ll M_2$ the model could  either favour the normal ordering case, $m_{\nu 3} > m_{\nu 2} > m_{\nu 1}$, or the inverted one, $m_{\nu 2} > m_{\nu 3} > m_{\nu 1}$, respectively.

We have performed a numerical analysis and found that in order to successfully accommodate the experimental values of the leptonic mixing angles, the leptonic CP violating phase and the neutrino mass squared splittings, soft-breaking Majorana terms need to be introduced. The correlations of the effective mass parameter of neutrinoless double beta decay with the lightest neutrino mass and leptonic CP phase for the scenario of normal neutrino mass hierarchy are shown in Figure \ref{Leptoncorrelations}.

\section{Implications on FCNCs}
\label{sec:pheno}
\noindent
Another aspect to be discussed is the amount of tree-level FCNCs. To explicitly show their smallness we consider the best-fit point values coming from the fit to the masses and mixing. 

In the quark sector, the Yukawa matrices parametrising the coupling to each Higgs doublet $\{ \Phi_1, \Phi_2, \Phi_S \}$ are denoted by ${\bf \Gamma}_1, {\bf \Gamma}_2, {\bf \Gamma}_S$, respectively. In the fermionic mass basis, they take the explicit form,
\begin{equation} 
\begin{gathered}
\widetilde{\bf \Gamma}_1^u  \sim  \begin{pmatrix}
 10^{-7} & 10^{-4} & 10^{-7} \\
 10^{-6} & 10^{-3} & 10^{-5} \\
 10^{-2} & 10^{-5}& 1
\end{pmatrix} , \quad
\widetilde{\bf \Gamma}_2^u  \sim  \begin{pmatrix}
 10^{-6} & 10^{-3} & 10^{-8} \\
 10^{-5} & 10^{-4} & 10^{-7} \\
 1 & 10^{-4} & 10^{-2}
\end{pmatrix} , \\
\widetilde{\bf \Gamma}_S^u  \sim  \left(
\begin{array}{ccc}
 10^{-4} & 10^{-7} & 10^{-5} \\
 10^{-5} & 10^{-8} & 10^{-4} \\
 0 & 0 & 10^{-9} \\
\end{array}
\right) ,
\\
\widetilde{\bf \Gamma}_1^d  \sim  \begin{pmatrix}
 10^{-5} & 10^{-4} & 10^{-5} \\
 10^{-4} & 10^{-4} & 10^{-4} \\
 10^{-4} & 10^{-3} & 10^{-3} 
\end{pmatrix}
+ i
 \begin{pmatrix}
  10^{-6} & 10^{-5}  & 10^{-4}\\
  10^{-5} &  10^{-6} & 10^{-4} \\
 10^{-3} & 10^{-3}&  10^{-4}
\end{pmatrix} , 
\\
\widetilde{\bf \Gamma}_2^d  \sim  
\begin{pmatrix}
 10^{-4} & 10^{-4} & 10^{-4} \\
 10^{-4} & 10^{-4} & 10^{-4} \\
 10^{-3} & 10^{-3} & 10^{-4} 
\end{pmatrix}
+ i
 \begin{pmatrix}
  10^{-4} & 10^{-4}  & 10^{-4}\\
  10^{-4} &  10^{-4} & 10^{-4} \\
 10^{-3} & 10^{-4}&  10^{-6}
\end{pmatrix} , 
\end{gathered}
\label{scalarfermioncouplings1}    
\end{equation}
\begin{equation}
\begin{gathered}
\widetilde{\bf \Gamma}_S^d  \sim  \begin{pmatrix}
 10^{-3} & 10^{-2} & 10^{-3} \\
 10^{-2} & 10^{-2} & 10^{-2} \\
 10^{-2} & 10^{-1} & 1 
\end{pmatrix}
+ i
 \begin{pmatrix}
  10^{-4} & 10^{-3}  & 10^{-2}\\
  10^{-3} &  10^{-3} & 10^{-2} \\
 10^{-1} & 10^{-1}&  10^{-2}
\end{pmatrix} ,
\end{gathered}
\label{scalarfermioncouplings2}    
\end{equation}
where we have also employed $v_1 \simeq 174 \text{ GeV}$,  $v_2 \simeq 8.14 \text{ GeV}$,
and $v_s \simeq 2.63 \text{ GeV}$. 

Now, in the scalar sector, we consider an explicit realization of the previously discussed benchmark scenario. There we assume
\begin{equation} { 
\begin{gathered}
|\mu_D| = 88.72 \text{ GeV} ,  \qquad |\mu_{12}| = 76  \text{ GeV} , \\
|\mu_S| = 1.5  \text{ TeV}, \qquad \lambda_5 = 0.13 \;,
\end{gathered}   }
\end{equation}
and through Eq.~\eqref{eq:VEVs}, $v_2 \simeq 8.14 \text{ GeV}$,
and $v_s \simeq 2.63 \text{ GeV}$ we obtain the explicit value for $\{\mu_{2},\mu_{S2} \}$.
This explicitly imply a degenerate scalar mass spectra:
\begin{equation} {
\begin{gathered}
m_h = 125 \text{ GeV} , \quad
m_{H_1^0} = 320 \text{ GeV} , \quad
m_{H_2^0} = 1.57 \text{ TeV}, \\
m_{A_1^0} = m_{H^\pm_1} = 320 \text{ GeV} , \quad
m_{A_2^0} = m_{H^\pm_2} = 1.57 \text{ TeV} .
\end{gathered}    }
\end{equation}
It is interesting to also remark here an important feature of our model: although we have small VEVs the scalar masses are very heavy. This is due to the fact on how we softly-broke the flavour symmetry and went into inducing VEVs.

The orthogonal transformation bringing the scalar sector to its mass basis is
approximately characterized by the same matrix
\begin{equation} {
{\bf R} \simeq \begin{pmatrix}
-0.998897 & 0.044653 & 0.014531 \\
0.046952 & 0.954338 & 0.295015\\
-0.000694 & 0.295372 & -0.95538
\end{pmatrix} \;,    }
\end{equation}
where we follow the convention ${\bf R} {\bf m}^2 {\bf R}^T = {\bf m}^2_\text{diag}$.
Notice how the smallness of the off-diagonal elements in ${\bf R}$ imply additional suppression factors when computing FCNCs. That together with the fact that all the scalars, neutral and charged, have sufficiently heavy masses  guarantee that FCNCs should be well below their upper bounds. 

\begin{figure}[h]
	\resizebox{7.5cm}{5.5cm}{\vspace{-2cm}%
		\includegraphics{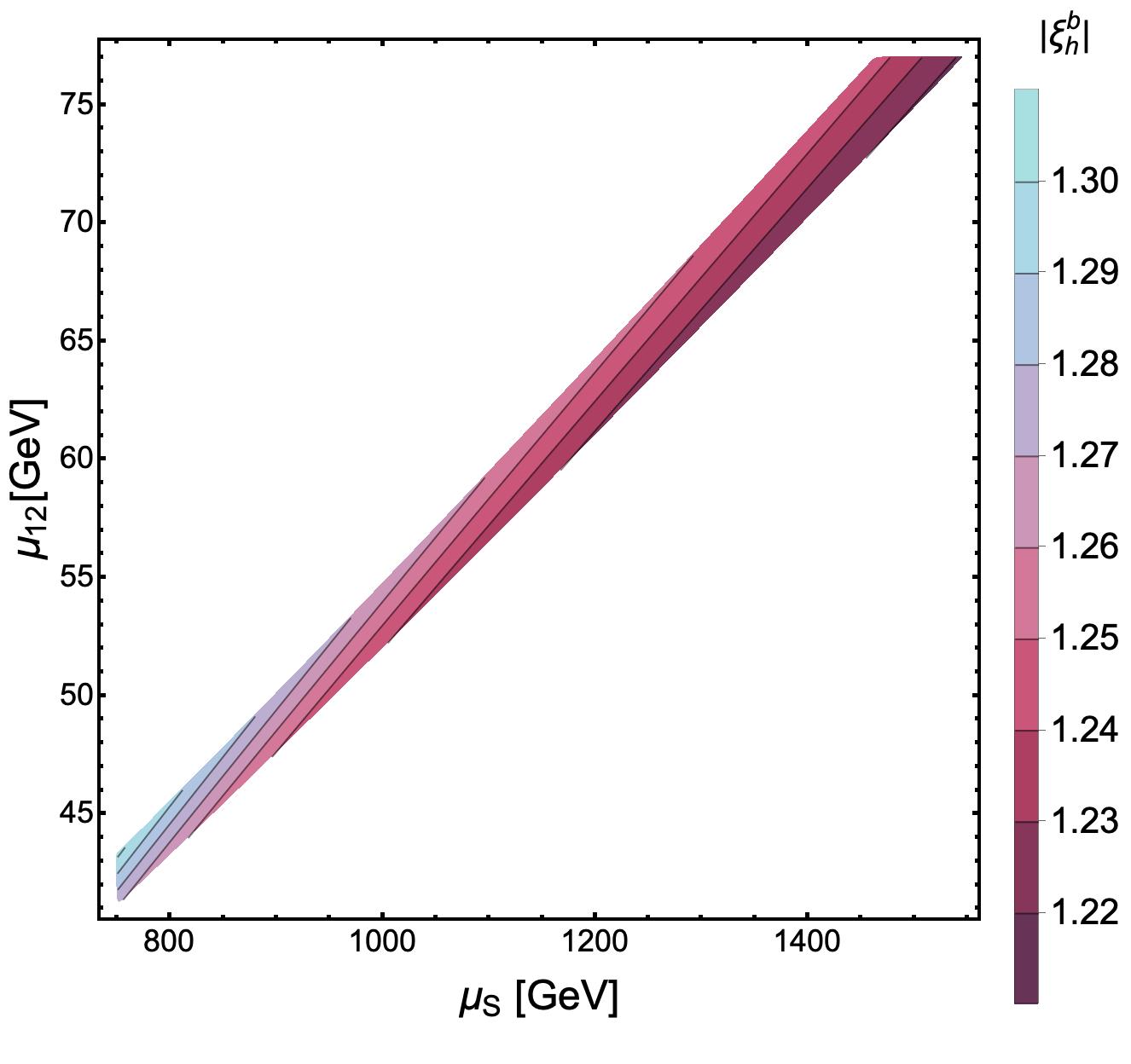}		
	}\vspace{0cm}
	\resizebox{7.5cm}{5.5cm}{\vspace{-2cm}%
		\includegraphics{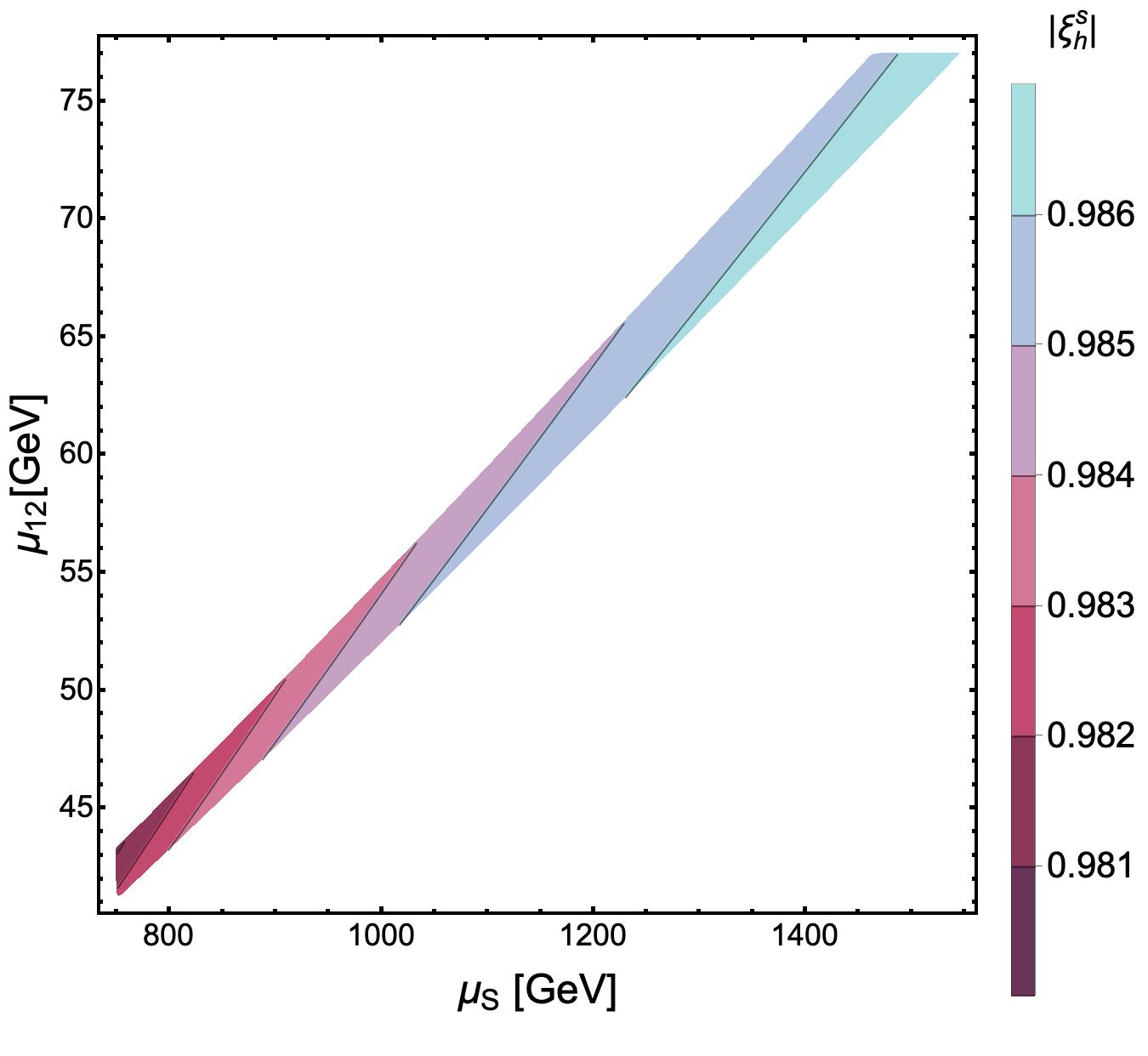}		
	}
	\caption{Effective couplings of the charm and bottom quarks to the SM-like Higgs.}
	\label{fig:Yukawacorrelations}
\end{figure}

On the other hand, we can now also investigate the effective couplings between the fermions and the SM-like Higgs:
\begin{equation}
-{\cal L}_Y \supset \sum_f \frac{m_f}{246 \text{ GeV}} \xi_h^f \bar{f} f h \;, 
\end{equation}
where 
\begin{equation}
\xi_h^{f_i} = \frac{\widetilde{\bf \Gamma}_1^{f,ii} {\bf R}_{11}+\widetilde{\bf \Gamma}_2^{f,ii}  {\bf R}_{21}+\widetilde{\bf \Gamma}_s^{f,ii} {\bf R}_{31}}{\widetilde{\bf \Gamma}_1^{f,ii} {\bf R}_{11}+\tan \beta\widetilde{\bf \Gamma}_2^{f,ii}  {\bf R}_{21}+\frac{v_s}{v_1}\widetilde{\bf \Gamma}_s^{f,ii} {\bf R}_{31}} \;.
\end{equation}
In our case, we find for { $\mu_{12} \in [35,77] \text{ GeV}$ and $\mu_S \in [750,2000] \text{ GeV}$ } the ranges: 
\begin{equation} {
\begin{gathered}
\xi^t_h = 1.00 \;, \quad
|\xi^b_h| \in [1.2,1.3] \;, \quad 
\xi^c_h = 1.00 
\\
|\xi^s_h| =0.98 \;, \quad
|\xi^d_h| \in [0.80,0.84] \;,
\end{gathered} }
\end{equation} 
which are in agreement to the most recent combined fits of data taken at $\sqrt{s} = 13 \text{ TeV}$~\cite{Sirunyan:2018koj}
\begin{align} 
    \kappa_t & = 1.02^{+0.19}_{-0.15} \;, \quad
    \; \, \kappa_\tau = 0.93^{+0.13}_{-0.13} \;, \\
    \kappa_b & = 0.91^{+0.17}_{-0.16} \;, \quad
    \; \kappa_\mu = 0.72^{+0.50}_{-0.72} \;.
\end{align}
while the lightest families to the latest global fits~\cite{Kagan:2014ila,Perez:2015aoa} with large upper bounds $\kappa_f < {\cal O}(10-100)$. Note how Figure~\ref{fig:Yukawacorrelations} shows a correlated behaviour between the strange and bottom quarks: the larger the one the smaller the other, for a discussion on the reach on projected sensitivities of future colliders see~\cite{deBlas:2019rxi}. We expect the analysis for the charged-leptons to be very similar to the down quarks, as they have similar fermion mass spectra and a similar structured mass matrix.

{
\section{Higgs diphoton decay rate}
\label{sec:photons}
\noindent
The decay rate for the $h\rightarrow \gamma \gamma$ process takes the form:
\begin{eqnarray}
\Gamma(h \rightarrow \gamma\gamma) &=& \dfrac{\alpha_{em}^2 m_h^3}{256 \pi^3 v^2}\left|\sum_f a_{hff} N_C Q_f^2 F_{1/2}(\rho_f)\notag\right.\\
&&\left.+a_{hWW} F_{1}(\rho_W)+\sum_{k=1,2}\frac{C_{hH^{\pm}_kH^{\mp}_k}v}{2m^{2}_{H^{\pm}_k}}F_{0}(\rho_{H^{\pm}_k})\right|^2,\notag\\
\end{eqnarray}
where $\rho_i$ are the mass ratios $\rho_i= \frac{m_h^2}{4 M_i^2}$ with $M_i=m_f, M_W$; $\alpha_{em}$ is the fine structure constant; $N_C$ is the color factor ($N_C=1$ for leptons and $N_C=3$ for quarks) and $Q_f$ is the electric charge of the fermion in the loop. From the fermion-loop contributions we only consider the dominant top quark term. Furthermore, $C_{hH^{\pm}_kH^{\mp}_k}$ is the trilinear coupling between the SM-like Higgs and a pair of charged Higges, whereas $a_{htt}$ and $a_{hWW}$ are the deviation factors from the SM Higgs–top quark coupling and the SM Higgs–W gauge boson coupling, respectively (in the SM these factors are unity). Such deviation factors are very close to unity in our model, which is a consequence of the numerical analysis of its scalar, Yukawa and gauge sectors.

Furthermore, $F_{1/2}(z)$ and $F_{1}(z)$ are the dimensionless loop factors for spin-$1/2$ and spin-$1$ particles running in the internal lines of the loops. They are given by:
\begin{align}
F_{1/2}(z) &= 2(z + (z -1)f(z))z^{-2}, \\
F_{1}(z) &= -2(2z^2 + 3z + 3(2z-1)f(z))z^{-2}, \\
F_{0}(z) &= -(z - f(z))z^{-2},
\end{align}
with
\begin{align}
f(z) = \left\{ \begin{array}{lcc}
\arcsin^2 \sqrt{2} & \text{for}  & z \leq 1 \\
\\ -\frac{1}{4}\left(\ln \left(\frac{1+\sqrt{1-z^{-1}}}{1-\sqrt{1-z^{-1}}-i\pi} \right)^2 \right) &  \text{for} & z > 1\\
\end{array}
\right.
\end{align}

In order to study the implications of our model in the decay of the $126$ GeV Higgs into a photon pair, one introduces the Higgs diphoton signal strength $R_{\gamma \gamma}$, which is defined as:
\begin{align}
R_{\gamma \gamma} = \frac{\sigma(pp \to h)\Gamma(h \to \gamma \gamma)}{\sigma(pp \to h)_{SM}\Gamma(h \to \gamma\gamma)_{SM}} \simeq a^{2}_{htt} \frac{\Gamma(h \to \gamma \gamma)}{\Gamma(h \to \gamma \gamma)_{SM}}.
\label{eqn:hgg}
\end{align}
That Higgs diphoton signal strength, normalizes the $\gamma \gamma$ signal predicted by our model in relation to the one given by the SM.  Here we have used the fact that in our model, single Higgs production is also dominated by gluon fusion as in the Standard Model.

The ratio $R_{\gamma \gamma}$ has been measured by CMS and ATLAS collaborations with the best fit signals \cite{Sirunyan:2018ouh,Aad:2019mbh}: 
\begin{align}
R^{CMS}_{\gamma \gamma} = 1.18^{+0.17}_{-0.14} \quad \text{and} \quad
R^{ATLAS}_{\gamma \gamma} = 0.96 \pm 0.14.
\label{eqn:rgg}
\end{align}
The correlation of the Higgs diphoton signal strength with the charged scalar mass $m_{H^{\pm}_1}$ is shown in Figure \ref{Higgsdiphoton}, which indicates that our model successfully accommodates the current Higgs diphoton decay rate constraints. Furthermore, as indicated by Figure \ref{Higgsdiphoton}, our model favours a Higgs diphoton decay rate very close to the SM expectation.
\begin{figure}[th]
	\resizebox{7.5cm}{5cm}{\vspace{-2cm}%
		\includegraphics{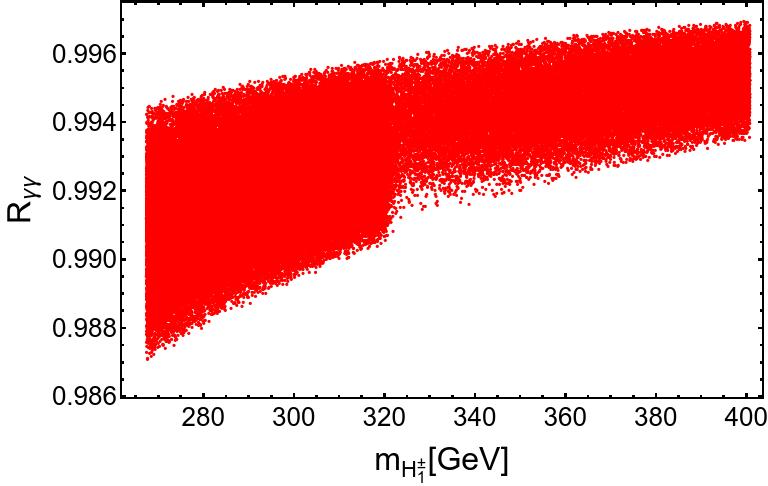}		
	}
	\caption{Correlation of the Higgs diphoton signal strength with the charged scalar mass $m_{H^{\pm}_1}$.}
	\label{Higgsdiphoton}
\end{figure}

\section{Heavy scalar production at a proton-proton collider}
\label{sec:HeavyScalar}
\noindent
This section deals with the discussion of the singly heavy scalar $H^0_{1}$ production at a proton-proton collider. Such production mechanism at the LHC is dominated by the gluon fusion mechanism, which is a one-loop process mediated by the top quark. Thus, the total $H^0_{1}$ production cross section in proton-proton collisions with center of mass energy $\sqrt{s}$ takes the form:
\begin{eqnarray}
\sigma _{pp\rightarrow gg\rightarrow H^0_{1}}\left( s\right) &=&\frac{\alpha
	_{S}^{2}a^2_{H^0_{1}t\bar{t}}m_{H^0_{1}}^{2}}{64\pi v^{2}S}\left[I\left( \frac{m_{H^0_{1}}^{2}}{m_{t}^{2}}%
\right)\right]^{2}\notag \\
&&\times \int_{\ln \sqrt{\frac{m_{H^0_{1}}^{2}}{S}}}^{-\ln \sqrt{\frac{%
			m_{H^0_{1}}^{2}}{s}}}f_{p/g}\left( \sqrt{\frac{m_{H^0_{1}}^{2}}{s}}e^{y},\mu
^{2}\right)\notag\\
&&\times f_{p/g}\left( \sqrt{\frac{m_{H^0_{1}}^{2}}{s}}e^{-y},\mu
^{2}\right) dy,
\end{eqnarray}
where $f_{p/g}\left( x_1,\mu ^2 \right) $ and $f_{p/g}\left(x_2,\mu
^2 \right) $ are the distributions of gluons in the proton which carry
momentum fractions $x_1$ and $x_2$ of the proton, respectively. Furthermore $%
\mu =m_{H_{1}}$ is the factorization scale, whereas $I(z)$ has the form:
\begin{equation}
I(z)=\int_{0}^{1}dx\int_{0}^{1-x}dy\frac{1-4xy}{1-zxy}.
\label{g1a}
\end{equation}
\begin{figure}[h]
	\resizebox{7.5cm}{6cm}{\vspace{-2cm}%
		\includegraphics{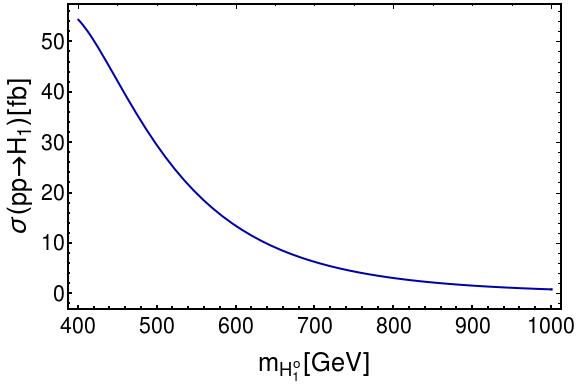}		
	}
	\caption{Total cross section for the $H_1^0$ production via gluon fusion
		mechanism at the LHC for $\protect\sqrt{s}=13$ TeV and as a function of the heavy scalar mass $m_{H^0_{1}}$.}
	\label{ggtoH1}
\end{figure}
\begin{figure}[h]
	\resizebox{7.5cm}{6cm}{\vspace{-2cm}%
		\includegraphics{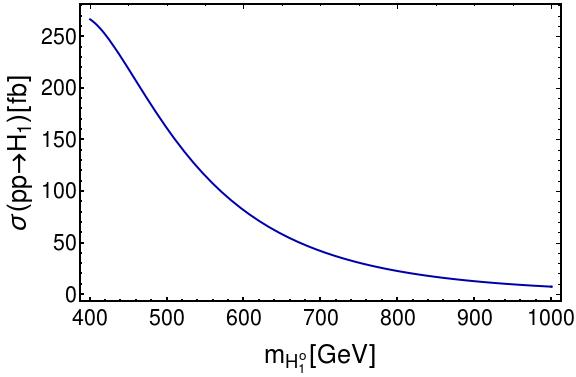}		
	}
	\caption{Total cross section for the $H_1^0$ production via gluon fusion
		mechanism at the proposed energy upgrade of the LHC with $\protect\sqrt{s}=28$ TeV and as a function of the heavy scalar mass $m_{H^0_{1}}$.}
	\label{ggtoH1for28TeV}
\end{figure}
\begin{figure}%[H]
	\resizebox{7.5cm}{6cm}{\vspace{-2cm}%
		\includegraphics{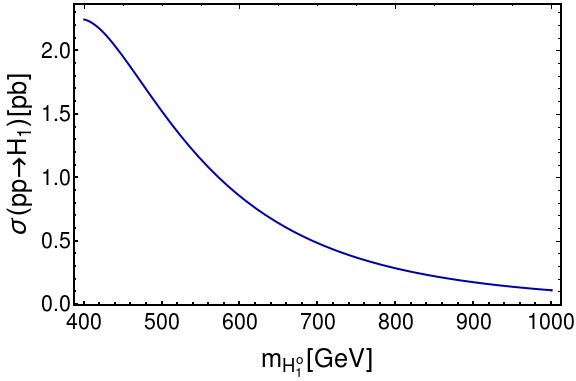}		
	}
	\caption{Total cross section for the $H_1^0$ production via gluon fusion
		mechanism at $\protect\sqrt{s}=100$ TeV proton-proton collider and as a function of the heavy scalar mass $m_{H^0_{1}}$.}
	\label{ggtoH1for100TeV}
\end{figure}
Figure~\ref{ggtoH1} shows the $H^0_{1}$ total production cross section at the
LHC via gluon fusion mechanism for $\sqrt{s}=13$ TeV, as a function of the scalar mass $m_{H^0_{1}}$, which is taken to
range from $400$ GeV up to $1$ TeV. Furthermore, the coupling $a_{H^0_{1}t\bar{t}}$ of the heavy scalar $H^0_{1}$ with the top-antitop pair has been set to be equal to $0.1$, which is consistent with Eqs.~\eqref{scalarfermioncouplings1} and~\eqref{scalarfermioncouplings2}. In the aforementioned region of masses for the heavy $H_1$ scalar, we find that the total production cross section ranges from $54$ fb up to $0.9$ fb. However, at the proposed energy upgrade of the LHC with $\protect\sqrt{s}=28$ TeV, the total cross section for the $H^0_{1}$ is enhanced reaching values between  $267$ fb and $8$ fb in the aforementioned mass range as shown in Figure~\ref{ggtoH1for28TeV}. Besides that, $H^0_{1}$ total production cross section is significantly enhanced at a $\sqrt{s}=100$ TeV proton-proton collider, since it takes values ranging from $2.2$ pb up to $0.1$ pb in the same region of masses, as shown in Figure~\ref{ggtoH1for100TeV}. Finally, in the scenario where $H^0_{1}$ is the lightest among the heavy scalars, it can decay, for instance into down type quark-antiquark, charged lepton-antilepton pairs thus leading to multijet and (or) multilepton final states. Given that the parameter space considered in this work is very close to the decoupling limit~\cite{Haber:1994mt,Gunion:2002zf}, 
the pair production of the $H_1^0$  scalar will have a small impact in the multilepton or multijet production over the SM expectation.
}

\section{Discussion}
\label{sec:discussion}
\noindent
The main feature of the presented model is that, when $\tan \beta \rightarrow 0$, all the quark mixing becomes trivial while in the lepton sector only the solar and 
reactor mixing angles follow the same fate.  
The atmospheric one is the only non-zero mixing angle. However, it does receive
contributions that could account for 15$\%$ of its experimental value.
Furthermore, in the same limit, the complete first generation of fermions become massless. 
Hence, all mixing angles are proportional, or at least related, to $\tan \beta$. This is the main result of our discussion. 

Now, it is important to question if this result could have been 
achieved with other symmetry groups. To investigate it we need to realize
that Eq.~\eqref{eq:DasRelation} is a direct consequence of the mass matrices
having the form
\begin{align} \label{eq:condition}
{\bf M} = y_1 \begin{pmatrix}
0 & 0 & v_2 \\
0 & 0 & v_1 \\
0 & 0 & 0
\end{pmatrix} + y_2 
\begin{pmatrix}
0 & 0 & 0 \\
0 & 0 & 0 \\
v_2 & v_1 & 0
\end{pmatrix} \; ,
\end{align}
up to some possible minus signs; this kind of matrix is 
then diagonalised by Eq.~\eqref{eq:rotation}. This structure points
to those symmetry groups which have doublets in their irreducible
representations. We are mainly interested in those types of groups where
the tensorial product of two doublets contains: i) at least one 
doublet with two singlets or ii) no doublets but four singlets. The latter
are characterised by $\mathbb{D}_{2n}$, $\mathbb{Q}_n$, ${\Sigma}_n$, etc.~\cite{Ishimori:2010au}. 
In the case of the former possibility, to generate Eq.~\eqref{eq:condition} we still
require an additional auxiliary symmetry, e.g. $\mathbb{Z}_k$, to forbid
the extra terms\footnote{Take for example $\mathbb{S}_3$. It is possible to show that with 
$\{Q_{3L}, u_{3R} \}\sim {\bf 1}$, $d_{3R}\sim {\bf 1}'$, $\{\Phi_1,\Phi_2\} \sim {\bf 2}$, $\{Q_{1L},Q_{2L}\}\sim {\bf 2}$
$\{d_{1R},d_{2R}\} \sim {\bf 2}$, and $\{ u_{1R},u_{2R}\} \sim {\bf 2}$ the relation is not possible unless 
one invokes a $\mathbb{Z}_2$ that differentiates flavour doublets (-) from flavour singlets (+) such
that the direct product of three doublets is forbidden. Moreover, realize that $\mathbb{Z}_2 \times \mathbb{S}_3 \simeq \mathbb{D}_6$.}. Notice that among all
the different choices, the most minimal is the one realized here with $\mathbb{D}_4$. 

\section{Conclusions}
\label{sec:conclusions}
\noindent
We have considered a generalization of the original idea given in Ref.~\cite{Das:2019itj} 
where the Cabibbo angle was expressed in terms of the physical parameter $\beta$
commonly appearing in 2HDMs, $\theta_c = 2 \beta$. The original proposal only had the
two heavy quark generations with non-zero masses and no mixing allowed with the third
generation. By adding a third Higgs doublet with a small VEV and assuming a conveniently large decoupling mass, we have been able to sufficiently perturb the original model and not only write all the quark mixing angles in terms of $\beta$ but also do the same in the lepton sector. 
Additionally, we have discussed how the suggested relation between these two angles ($\theta_c$ and $\beta$) is a consequence of symmetry groups with doublets in their irreducible representations satisfying tensorial products of the type: i) ${\bf 2}\times {\bf 2} \sim {\bf 1}+ {\bf 1}^\prime
+{\bf 2}$ or ii) ${\bf 2}\times {\bf 2} \sim {\bf 1}+ {\bf 1}^\prime + {\bf 1}^{''} + {\bf 1}^{'''}$,
where for the former case an auxiliary symmetry is still required to forbid extra terms not appearing in the latter case. { An important aspect, when relating $\beta$ to all the quark mixing, is that the Cabibbo angle cannot be fully explained by simply $\theta_c =2|\beta|$. It requires further corrections reaching { $\sim 60 \%$}.} {Besides that, we have performed a numerical analysis of the scalar sector finding linear correlations among the non SM scalar masses. Furthermore, we have studied the singly scalar $H^0_{1}$ production at proton-proton collider via gluon fusion mechanism at $\sqrt{s}=13$ TeV, $\sqrt{s}=28$ TeV and $\sqrt{s}=100$ TeV obtaining total cross sections close to about $50$ fb, $270$ fb and $2$ pb, respectively for a $400$ GeV heavy scalar mass. In addition, we have also shown that our model successfully accommodates the current Higgs diphoton decay rate constraints, yielding a Higgs diphoton decay rate very close to the SM expectation.} Finally, although this framework contains tree-level FCNCs { we have shown} they are sufficiently suppressed for heavy scalar masses. { We have left for future work the study on how small could the scalar masses be without creating dangerous FCNCs.}

\section*{Acknowledgments}
\noindent
The visit of U.J.S.S. to Universidad Técnica Federico Santa María (UTFSM) was
supported by Chilean Grant Fondecyt No. 1170803.
U.J.S.S. would like to thank the members of 
the UTFSM-group in Valpara\'iso for their hospitality during his visit, where 
part of this work was done.
U.J.S.S. acknowledges support by CONACYT (M\'{e}xico). A.E.C.H. and C.O.D. acknowledge support from FONDECYT (Chile) grants 1170803 and 1170171, and from CONICYT PIA/Basal FB0821.

{ We would also like to acknowledge the anonymous referee whose critics and observations have strongly improved the quality of the present paper during its revision.}

\appendix

\section{THE PRODUCT RULES FOR $\mathbb{D}_4$}
\label{app:rules}
\noindent {
The group $\mathbb{D}_4$ has two generators, ${\bf a}$ and ${\bf b}$, such that ${\bf a}^4 = {\bf b}^2 =I$. Also, it has five irreducible representations: one 2-dimensional and four 1-dimensional. The latter are characterized by the eigenvalues of the generators ${\bf a}$ and ${\bf b}$ respectively as $1_{++}$, $1_{+-}$, $1_{-+}$ and $1_{--}$. In this work we use the convention where the $\mathbb{D}_4$ singlets are denoted as $1_{b,ab}$, as done in \cite{Ishimori:2010au}. That convention implies that for $1_{-+}$, $b=-1$, $ab=1$, which corresponds to $b=-1$.

For the 2-dimensional representation, we choose the following basis for the generators: 
\[
a= 
\left( 
\begin{array}{c c}
i & 0
\\
0 & -i
\end{array}
\right) ,
\qquad 
b= 
\left( 
\begin{array}{c c}
0& 1
\\
1 & 0
\end{array}
\right)
\]

The multiplication rule of the singlets is simply given by 
\begin{align}
\mathbf{1}_{x_1 y_1} \times \mathbf{1}_{x_2 y_2} = \mathbf{1}_{x_3 y_3}
\end{align}
where $x_3 = x_1 x_2$ and $y_3 = y_1 y_2$.

Now, consider two doublets that transform under $\mathbb{D}_4$: 
\[
\Phi = 
\left(
\begin{array}{c}
\phi_1 \cr \phi_2
\end{array}
\right) ,
 \qquad 
\Psi = 
\left(
\begin{array}{c}
\psi_1 \cr \psi_2
\end{array}
\right)
\]
The product of two 2-d representations $\Phi \times \Psi$, or $\Phi^\ast \times \Psi$, can be decomposed in the following 1-d representations:

\begin{align}
1_{++}: & \qquad \Phi^T \sigma_1 \Psi ; \hspace{18pt} \Phi^\dagger \Psi ,
\\
1_{--}: & \qquad \Phi^T \epsilon \Psi ; \hspace{24pt} \Phi^\dagger \sigma_3 \Psi ,
\\
1_{+-}: & \qquad \Phi^T  \Psi ; \hspace{28pt} \Phi^\dagger \sigma_1 \Psi ,
\\
1_{-+}: & \qquad \Phi^T \sigma_3 \Psi ; \hspace{16pt} \Phi^\dagger \epsilon \Psi .
\end{align}

Here $\epsilon = i\sigma_2$, while $\sigma_1,\sigma_2, \sigma_3$ are the well-known Pauli matrices. More explicitly:
\begin{itemize}
\item 
$(\phi_1 \psi_2 + \phi_2 \psi_1) $ and $(\phi_1^\ast \psi_1 + \phi_2^\ast  \psi_2)$ transform as $1_{++}$,

\item 
$(\phi_1 \psi_2 - \phi_2 \psi_1) $ and $(\phi_1^\ast \psi_1 - \phi_2^\ast  \psi_2)$ transform as $1_{--}$,

\item 
$(\phi_1 \psi_1 + \phi_2 \psi_2) $ and $(\phi_1^\ast \psi_2 + \phi_2^\ast  \psi_1)$ transform as $1_{+-}$,

\item 
$(\phi_1 \psi_1 - \phi_2 \psi_2) $ and $(\phi_1^\ast \psi_2 - \phi_2^\ast  \psi_1)$ transform as $1_{-+}$.

\end{itemize}

Consequently, if the product between the two doublets does not involve complex conjugation then one simply has:
\begin{align} \label{eq:TensorialProductsD4}
\begin{split}
\mathbf{k} \times \mathbf{g} = & \, (k_1 g_2 + k_2 g_1)_{\mathbf{1}_{++}} + (k_1
g_2 - k_2 g_1)_{\mathbf{1}_{--}} 
\\ &
+ (k_1 g_1 + k_2 g_2)_{\mathbf{1}_{+-}} + (k_1 g_1 - k_2 g_2)_{\mathbf{1}_{-+}} \;,
\end{split}
\end{align}
{Furthermore, due to our choice of basis where one of the generators of the two-dimensional $\mathbb{D}_4$ representations is complex, the combination $\sigma_2 {\bf k}^*$  transforms as a $\mathbb{D}_4$ doublet, instead of ${\bf k}^*$, thus implying the following relation:
\begin{align} \label{eq:TensorialProductsD4complexdoublet}
\begin{split}
\mathbf{k}^*\times \mathbf{g} = & \, (k^*_2 g_2 + k^*_1 g_1)_{\mathbf{1}_{++}} + (k^*_2
g_2 - k^*_1 g_1)_{\mathbf{1}_{--}} 
\\ &
+ (k^*_2 g_1 + k^*_1 g_2)_{\mathbf{1}_{+-}} + (k^*_2 g_1 - k^*_1 g_2)_{\mathbf{1}_{-+}} \;,
\end{split}
\end{align}}
whereas for the singlet times doublet tensorial products we have:
\begin{align}
\begin{split}
(q)_{\mathbf{1}_{++}} \times \mathbf{g} = \begin{pmatrix}
 q g_1 \\
 q g_2
\end{pmatrix} \;,
 \qquad
(q)_{\mathbf{1}_{--}} \times \mathbf{g} = \begin{pmatrix}
 q g_1 \\
 -q g_2
\end{pmatrix}\;, \\
(q)_{\mathbf{1}_{+-}} \times \mathbf{g} = \begin{pmatrix}
 q g_2 \\
 q g_1
\end{pmatrix} \;,
 \qquad
(q)_{\mathbf{1}_{-+}} \times \mathbf{g} = \begin{pmatrix}
 q g_2 \\
 -q g_1
\end{pmatrix} \;.
\end{split}
\end{align}
{\begin{align}
\begin{split}
(q)_{\mathbf{1}_{++}} \times \mathbf{k^*} = \begin{pmatrix}
 q k^*_2 \\
 q k^*_1
\end{pmatrix} \;,
 \qquad
(q)_{\mathbf{1}_{--}} \times \mathbf{k^*} = \begin{pmatrix}
 q k^*_2 \\
 -q k^*_1
\end{pmatrix}\;, \\
(q)_{\mathbf{1}_{+-}} \times \mathbf{k^*} = \begin{pmatrix}
 q k^*_1 \\
 q k^*_2
\end{pmatrix} \;,
 \qquad
(q)_{\mathbf{1}_{-+}} \times \mathbf{k^*} = \begin{pmatrix}
 q k^*_1 \\
 -q k^*_2
\end{pmatrix} \;.
\end{split}
\end{align}}
where  we have denoted the doublets by $\mathbf{k} = (k_1,k_2)^T$ and $\mathbf{g} = (g_1 ,g_2)^T$.
}

\section{PHYSICAL COMPLEX PHASES}
\label{app:complexphases}
\noindent 
The following phase field redefinitions
\begin{equation}
\begin{gathered}
{\bf M}^\prime_u = {\bf \Lambda}_L^Q {\bf M}_u {\bf \Lambda}_R^u{}^{-1} \;, \qquad
{\bf M}^\prime_d = {\bf \Lambda}_L^Q {\bf M}_u {\bf \Lambda}_R^d{}^{-1} \;, \\
{\bf M}^\prime_e = {\bf \Lambda}_L^\ell {\bf M}_e {\bf \Lambda}_R^e{}^{-1}
\end{gathered}
\end{equation}
where ${\bf \Lambda}_{L,R} =\text{Diag}(e^{-i \alpha_{L(R)1}},e^{-i \alpha_{L(R)2}},e^{-i \alpha_{L(R)3}})$ and
with
\begin{equation}
\begin{gathered}
\alpha_{L1}^Q = \text{arg}(y^u_2) + \alpha_{R3}^u \;, \quad
\alpha_{L2}^Q = \text{arg}(y^u_2) + \alpha_{R3}^u \;, \\
\alpha_{L3}^Q = \text{arg}(y^u_3) +\text{arg}(y^u_2) - \text{arg}(y^u_1) + \alpha_{R3}^u \\
\alpha_{R1}^u = \alpha_{R2}^u = \text{arg}(y^u_2) - \text{arg}(y^u_1) +\alpha_{R3}^u \;, \\
\alpha_{R1}^d = \alpha_{R2}^d =  \text{arg}(y^u_2) - \text{arg}(y^d_1) +\alpha_{R3}^u \;, \\
\alpha_{R3}^d = \text{arg}(y^u_2) - \text{arg}(y^d_2) +\alpha_{R3}^u \;, 
\\
\alpha_{R1}^e = \alpha_{R2}^e = - \text{arg}(y_3^e) + \alpha_{L3}^\ell \;, \\
\alpha_{L1}^\ell = \alpha_{L2}^\ell = \text{arg}(y_1^e) - \text{arg}(y_3^e) + \alpha_{L3}^\ell \;, \\
\alpha_{R3}^e = \text{arg}(y_1^e) - \text{arg}(y_2^e) - \text{arg}(y_3^e) + \alpha_{L3}^\ell \;,\\
\end{gathered}
\end{equation}
can bring all the up quarks, down quarks, and charged leptons complex phases to
\begin{equation}
\begin{gathered}
y_1^u = 0 \;, \quad
y_2^u = 0;, \quad
y_3^u =0 \;, \\
y_1^d =0 \;, \quad y^d_2 = 0 \;, \quad y^e_1 = 0 \;, \\
y_2^e = 0 \;, \quad y_3^e = 0 \;, 
\end{gathered}
\end{equation}
while only remain as non-zero { the subset}: 
\begin{align}
y_3^d, y_4^d, y^e_4 \neq 0 \;.
\end{align}

\newpage

\section{SCALAR MASS MATRICES}
\label{app:scalar}
\noindent
{From the scalar potential, we find that the squared mass matrices for the CP even neutral, CP odd neutral, and charged scalar sectors are respectively given by:	}
\begin{widetext} {
\begin{align}
\begin{split}
{\bf m}^2_{CP-\text{even}}=&\resizebox{0.9\hsize}{!}{{\fontsize{3}{3}{\tiny$
\left(
\begin{array}{ccc}
2v_1^2 \left(\lambda _4+\lambda _5\right)+{\tan\beta}\left(- v_s^2 \lambda _7 + \mu_{12}^2\right)
& 
-\mu _{12}^2 +  2 v_1 v_2 \left(2 \lambda _2+\lambda _4-\lambda _5\right) +  \lambda_7 v_s^2
& 
 v_s v_1 \left(\lambda _6+\lambda _8\right) + 2 v_s v_2 \lambda_7
\\
-\mu _{12}^2 +  2 v_1 v_2 \left(2 \lambda _2+\lambda _4-\lambda _5\right) +  \lambda_7 v_s^2
&
2v_2^2 \left(\lambda _4+\lambda _5\right)+{\cot\beta}\left(-\lambda_7 v_s^2  + \mu _{12}^2\right) +\dfrac{v_s}{2 v_2}\mu_{S2}^2 
&
 2 v_1 v_s \lambda_7+v_2 v_s (\lambda_6 + \lambda_8)-\frac{\mu _{\text{S2}}^2}{2}
\\
 v_s v_1 \left(\lambda _6+\lambda _8\right) + 2 v_s v_2 \lambda_7
& 
2 v_1 v_s \lambda_7+v_2 v_s (\lambda_6 + \lambda_8)-\frac{\mu _{\text{S2}}^2}{2}
& 
2 \lambda _1 v_s^2+\frac{v_2 }{2v_s} \mu _{\text{S2}}^2 \\
\end{array}
\right)$}}}
\end{split} \;,
\end{align}
\begin{align}
\begin{split}
{\bf m}^2_{CP-\text{odd}}=&\resizebox{0.9\hsize}{!}{{\fontsize{3}{3}{\tiny$
\left(
\begin{array}{ccc}
-2 v_2^2 (\lambda_2 + \lambda_3) - {\tan \beta} \left[\mu_{12}^2 - \lambda_7 v_s^2 \right]
& 
-\mu_{12}^2 + 2\left(\lambda _2+\lambda _3\right) v_1 v_2 -\lambda_7 v_s^2
& 
2 \lambda _7 v_2 v_s 
\\
-\mu_{12}^2 + 2\left(\lambda _2+\lambda _3\right) v_1 v_2 -\lambda_7 v_s^2
& 
-2 v_1^2 (\lambda_2 + \lambda_3) + {\cot \beta} \left( \mu_{12}^2 -  \lambda_7 v_s^2 \right)
+\frac{v_s}{2 v_2} \mu_{S2}^2
& 
2\lambda _7 v_1 v_s - \frac{\mu _{\text{S2}}^2}{2}
\\
2\lambda _7 v_2 v_s 
& 
2\lambda _7 v_1 v_s - \frac{\mu _{\text{S2}}^2}{2}
& 
-4 \lambda_7 v_1 v_2 + \frac{v_2}{2 v_s}\mu_{S2}^2 \\
\end{array}
\right)$}}}
\end{split} \;,
\end{align}
\begin{align}
\begin{split}
{\bf m}^2_{\text{charged}}= &\resizebox{0.9\hsize}{!}{{\fontsize{3}{3}{\tiny$
\left(
\begin{array}{ccc}
-2 v_2^2 \lambda_2  - \tan \beta \left( - \mu_{12}^2 + \lambda_7 v_s^2  \right) - \frac{\lambda_6}{2} v_s^2 
& 
-\mu_{12}^2 + 2\lambda _2 v_1 v_2 
& 
\frac{1}{2} \lambda_6  v_1 v_s +  \lambda_7 v_2 v_s 
\\
-\mu_{12}^2 + 2\lambda _2 v_1 v_2 
& 
-2 v_1^2 \lambda_2  +  {\cot \beta} \left( \mu_{12}^2 -  \lambda_7 v_s^2 \right) + \frac{v_s}{2v_2} \left(-\lambda_6 v_2 v_s +  \mu_{S2}^2 \right)
& 
\frac{1}{2} (\lambda_6  v_2 v_s + 2 v_1 v_s \lambda_7 -\mu _{\text{S2}}^2
\\
\frac{1}{2} \lambda_6  v_1 v_s +  \lambda_7 v_2 v_s 
& 
\frac{1}{2} (\lambda_6  v_2 v_s + 2 v_1 v_s \lambda_7 -\mu _{\text{S2}}^2
& 
-\frac{1}{2}\left[ (v_1^2 + v_2^2)\lambda_6 + 4 v_1 v_2 \lambda_7 \right]  + \frac{v_2}{2v_s}\mu_{S2}^2 \\
\end{array}
\right)$}}}			 
\end{split} \;.
\end{align}    }
\end{widetext} 
\hspace{20cm}\newline
\hspace{20cm}\newline
\hspace{20cm}\newline
\hspace{20cm}\newline
\hspace{20cm}\newline
\hspace{20cm}\newline
%\hspace{20cm}\newline
%\hspace{20cm}\newline
%\newpage

\section{ALTERNATIVE SCENARIOS}
\label{app:scenarios}
\noindent
Let us consider the $\mathbb{D}_4$ assignments to a generic fermion sector as
\begin{align}
\begin{split} \psi_D \; =
\begin{pmatrix}
\psi_1 \\
\psi_2
\end{pmatrix} \sim {\bf 2} \;, \qquad (\psi = \psi_L,\psi_R) \;, \\
\psi_{3L} \sim {\bf 1}_{x} \quad \text{and} \quad \psi_{3R} \sim {\bf 1}_y \;,
\end{split}
\end{align}
where $\{{\bf 1}_x,{\bf 1}_y\} = \{ {\bf 1}_{++},{\bf 1}_{--},{\bf 1}_{+-},{\bf 1}_{-+} \}$ and $x$ and $y$ are not necessarily the same. Also let us fix the scalar sector to
\begin{align} \Phi_D =
\begin{pmatrix}
\Phi_1 \\
\Phi_2
\end{pmatrix} \sim {\bf 2} \;, \quad \text{and} \quad \Phi_S \sim {\bf 1}_z
\end{align}
where similarly for ${\bf 1}_z$ one has four different choices, $\{ {\bf 1}_{++},{\bf 1}_{--},{\bf 1}_{+-},{\bf 1}_{-+} \}$.

The basic \textit{lego} pieces in this kind of setup are{
\begin{align} 
[\overline{\psi}_{DL} \Phi_D] \psi_{3R} & =
\begin{cases}
(\overline{\psi}_{2L} \Phi_{2} \pm \overline{\psi}_{1L} \Phi_{1}) \psi_{3R}, \qquad {\bf 1}_{++,--} \\
(\overline{\psi}_{2L} \Phi_{1} \pm \overline{\psi}_{1L} \Phi_2) \psi_{3R}, \qquad {\bf 1}_{+-,-+} 
\end{cases} \\
[\overline{\psi}_{DL} \widetilde{\Phi}_D] \psi_{3R} & =
\begin{cases}
(\overline{\psi}_{2L} \widetilde{\Phi}_{1} \pm \overline{\psi}_{1L} \widetilde{\Phi}_{2}) \psi_{3R}, \qquad {\bf 1}_{++,--} \\
(\overline{\psi}_{2L} \widetilde{\Phi}_{2} \pm \overline{\psi}_{1L} \widetilde{\Phi}_1) \psi_{3R}, \qquad {\bf 1}_{+-,-+} 
\end{cases} 
\\
\overline{\psi}_{3L} [\Phi_D \psi_{DR}] & =
\begin{cases}
\overline{\psi}_{3L} (\Phi_1 \psi_{2R} \pm \Phi_2 \psi_{1R}),   \qquad {\bf 1}_{++,--} \\
\overline{\psi}_{3L} (\Phi_1 \psi_{1R} \pm \Phi_2 \psi_{2R}),  \qquad {\bf 1}_{+-,-+} 
\end{cases}
\\
\overline{\psi}_{3L} [\widetilde{\Phi}_D \psi_{DR}] & =
\begin{cases}
\overline{\psi}_{3L} (\widetilde{\Phi}_2 \psi_{2R} \pm \widetilde{\Phi}_1 \psi_{1R}),   \qquad {\bf 1}_{++,--} \\
\overline{\psi}_{3L} (\widetilde{\Phi}_2 \psi_{1R} \pm \widetilde{\Phi}_1 \psi_{2R}),  \qquad {\bf 1}_{+-,-+} 
\end{cases} \\
[\overline{\psi}_{DL} \psi_{DR}] \Phi_S & =
\begin{cases}
(\overline{\psi}_{2L} \psi_{2R} \pm \overline{\psi}_{1L} \psi_{1R} ) \Phi_S, \quad \; {\bf 1}_{++,--}\\
(\overline{\psi}_{2L} \psi_{1R} \pm \overline{\psi}_{1L} \psi_{2R} ) \Phi_S, \quad \; {\bf 1}_{+-,-+}
\end{cases}
\\
\overline{\psi}_{3L} \Phi_S \psi_{3R} & =
\begin{cases}
0, \qquad \qquad \qquad \, {{\bf 1}_x {\bf 1}_y {\bf 1}_z \neq {\bf 1}_{++}} \\
\overline{\psi}_{3L} \Phi_S \psi_{3R}, \qquad {{\bf 1}_x {\bf 1}_y {\bf 1}_z = {\bf 1}_{++}}
\end{cases}
\end{align}  }
where each of the different cases depends upon the flavour singlet field shown in each term.
The latter also means one has at least 3 Yukawa parameters up to 4, in the maximum case, per fermion species.  

{ Within our context,} there are in total $4^3 = 64$ possible { singlet} assignment combinations. However, we may reduce them to the following Yukawa structures {
\begin{align}
[\overline{\psi}_{DL} \Phi_D] \psi_{3R} & :
\begin{cases}
\begin{pmatrix}
0& 0 & \pm v_1 \\
0& 0 & v_2 \\
0& 0&0
\end{pmatrix}, \; \qquad {\bf 1}_{++,--} \\ \\
\begin{pmatrix}
0 & 0 & \pm v_2 \\
0 & 0 & v_1 \\
0 & 0 & 0
\end{pmatrix}, \qquad {\bf 1}_{+-,-+} 
\end{cases} 
\end{align}
\begin{align}
\overline{\psi}_{3L} [\Phi_D \psi_{DR}] & :
\begin{cases}
\begin{pmatrix}
0& 0 & 0 \\
0& 0 & 0 \\
\pm v_2 & v_1 &0
\end{pmatrix},   \qquad {\bf 1}_{++,--} \\ \\
\begin{pmatrix}
0& 0 & 0 \\
0& 0 & 0 \\
v_1 & \pm v_2&0
\end{pmatrix},  \qquad {\bf 1}_{+-,-+} 
\end{cases} 
\end{align}
\begin{align}
[\overline{\psi}_{DL} \widetilde{\Phi}_D] \psi_{3R} & :
\begin{cases}
\begin{pmatrix}
0& 0 & \pm v_2 \\
0& 0 & v_1 \\
0& 0&0
\end{pmatrix}, \; \qquad {\bf 1}_{++,--} \\ \\
\begin{pmatrix}
0 & 0 & \pm v_1 \\
0 & 0 & v_2 \\
0 & 0 & 0
\end{pmatrix}, \qquad {\bf 1}_{+-,-+} 
\end{cases}
\end{align}
\begin{align}
\overline{\psi}_{3L} [\widetilde{\Phi}_D \psi_{DR}] & :
\begin{cases}
\begin{pmatrix}
0& 0 & 0 \\
0& 0 & 0 \\
\pm v_1 & v_2 &0
\end{pmatrix},   \qquad {\bf 1}_{++,--} \\ \\
\begin{pmatrix}
0& 0 & 0 \\
0& 0 & 0 \\
v_2 & \pm v_1 &0
\end{pmatrix},  \qquad {\bf 1}_{+-,-+} 
\end{cases} 
\end{align}
\begin{align}
[\overline{\psi}_{DL} \psi_{DR}] \Phi_S & :
\begin{cases}
\begin{pmatrix}
\pm v_s & 0 & 0 \\
0 & v_s & 0 \\
0 & 0 & 0
\end{pmatrix}, \qquad  {\bf 1}_{++,--}\\ \\
\begin{pmatrix}
0 & \pm v_s & 0 \\ 
 v_s & 0 & 0\\
0& 0& 0
\end{pmatrix}, \qquad  {\bf 1}_{+-,-+}
\end{cases}           
\end{align}
\begin{align}                                
\overline{\psi}_{3L} \Phi_S \psi_{3R} & :
\begin{cases}
\begin{pmatrix}
0& 0 & 0 \\
0& 0 & 0 \\
0& 0&0
\end{pmatrix}, \quad  \; {{\bf 1}_x {\bf 1}_y {\bf 1}_z \neq {\bf 1}_{++}} \\ \\
\begin{pmatrix}
0& 0 & 0 \\
0& 0 & 0 \\
0& 0& v_s
\end{pmatrix}, \quad {{\bf 1}_x {\bf 1}_y {\bf 1}_z = {\bf 1}_{++}}
\end{cases}
\end{align}}
{ where for $\widetilde{\Phi}_S$ there are not different structures compared to those from $\Phi_S$.}

There are in fact two types of four major structures. The first type { is denoted by $\{{\bf M}_A, {\bf M}_B, {\bf M}_C, {\bf M}_D \}$} { where}:  {
\begin{align}
{\bf M}_A & =  
\begin{pmatrix}
\eta_{S}y_1 v_s& 0 & \eta_{3R} y_2 v_1 \\
0&  y_1 v_s &  y_2 v_2 \\
\eta_{3L} y_3 v_2& y_3 v_1& y_4 v_s
\end{pmatrix} \;, \\
{\bf M}_{B} & =  
\begin{pmatrix}
\eta_{S}y_1 v_s& 0 & \eta_{3R} y_2 v_1 \\
0&  y_1 v_s &  y_2 v_2 \\
 y_3 v_1& \eta_{3L} y_3 v_2& y_4 v_s
\end{pmatrix} \;, \\
{\bf M}_C & =  
\begin{pmatrix}
\eta_{S}y_1 v_s& 0 & \eta_{3R} y_2 v_2 \\
0&  y_1 v_s &  y_2 v_1 \\
\eta_{3L} y_3 v_2& y_3 v_1& y_4 v_s
\end{pmatrix} \;, \\
{\bf M}_{D} & =  
\begin{pmatrix}
\eta_{S}y_1 v_s& 0 & \eta_{3R} y_2 v_2 \\
0&  y_1 v_s &  y_2 v_1 \\
 y_3 v_1& \eta_{3L} y_3 v_2& y_4 v_s
\end{pmatrix} \;,
\end{align} }
{ while} the second one  $\{\overline{\bf M}_A, \overline{\bf M}_B, \overline{\bf M}_C, \overline{\bf M}_D \}$ { is given as}: {
\begin{align}
\overline{\bf M}_A & =  
\begin{pmatrix}
0&\eta_{S}y_1 v_s & \eta_{3R} y_2 v_1 \\ 
 y_1 v_s & 0 &  y_2 v_2 \\
\eta_{3L} y_3 v_2& y_3 v_1& y_4 v_s
\end{pmatrix} \;, \\
\overline{\bf M}_{B} & =  
\begin{pmatrix}
0&\eta_{S}y_1 v_s & \eta_{3R} y_2 v_1 \\ 
 y_1 v_s & 0 &  y_2 v_2 \\
 y_3 v_1& \eta_{3L} y_3 v_2& y_4 v_s
\end{pmatrix} \;, \\
\overline{\bf M}_C & =  
\begin{pmatrix}
0&\eta_{S}y_1 v_s  & \eta_{3R} y_2 v_2 \\
 y_1 v_s & 0 &  y_2 v_1 \\
\eta_{3L} y_3 v_2& y_3 v_1& y_4 v_s
\end{pmatrix} \;, \\
\overline{\bf M}_{D} & =  
\begin{pmatrix}
0&\eta_{S}y_1 v_s & \eta_{3R} y_2 v_2 \\
y_1 v_s & 0 &  y_2 v_1 \\
 y_3 v_1& \eta_{3L} y_3 v_2& y_4 v_s
\end{pmatrix} \;,
\end{align} }
where $\eta_{3R, 3L, S} = \{ +,- \}$. { The eight aforementioned mass matrices correspond to the down-quark (charged lepton) sector. The respective ones, for the up-type quarks (Dirac neutrinos), can be simply obtained by replacing $v_1 \leftrightarrow v_2$. The latter is a consequence of the complex nature of the flavor generators.} All these cases are already considering the possibility of having $y_4 = 0$. Moreover, note how all the eight different matrices become rank 2 whenever $y_1 \rightarrow 0$, which is not the case when we set to zero any of the other Yukawa parameters { except for $y_4$ which if additionally $\eta_{3R} \eta_{3L} \eta_S = -1$ then one could also get rank 2 matrices}. This tells us that the lightest mass should be proportional to $y_1$. 

Now, there is one fact we want to make sure our model possesses 
\begin{align}
	\tan \beta = \frac{v_2}{v_1} \simeq \frac{\theta_c}{2} \;,
\end{align}
up to a possible minus sign and where $\theta_c \approx 0.22$ is the Cabibbo mixing angle. In return, we get that $v_2 \simeq \beta v_1$. Additionally, to ensure a suitably definition of the $\beta$ angle (as defined in 2HDMs) we need to consider $v_s = \beta^2 v_1$.
Hence, without any loss of generality
\begin{equation}
v_1^2 + v_2^2 + v_s^2 = v_1^2 + \beta^2 v_1^2 + \beta^4 v_s^2 \approx v_1^2 + v_2^2.
\end{equation}

For last, Yukawa couplings are not necessarily order 1 numbers, but, in fact, in most cases they could end up being very small $y \leqslant 10^{-2}$ (of course, the only exception is the top quark Yukawa).

\section{RELATING THE QUARK MIXING ANGLES TO $\beta$}
\label{app:Relating}
\noindent
We first define the Cabibbo basis as the basis where the mixing matrix has already the corresponding contribution,
\begin{equation}
{\bf V}_{0} = \begin{pmatrix}
	\cos 2\beta & -\sin 2\beta & 0 \\
	\sin 2\beta & \cos 2\beta & 0\\
	0 & 0 & 1
\end{pmatrix} \;.
\end{equation}

The hermitian quark mass matrices, $\widetilde{\bf H} = {\bf R} {\bf M} {\bf M}^\dagger {\bf R}^T$, then take the exact form,
\begin{widetext}
\begin{equation} \label{eq:Hu}
\begin{gathered} {
\widetilde{\bf H}_u   = 
\begin{pmatrix}
(y_1^u)^2 v_s^2 & 0 & -2\frac{v_1 v_2 v_s y_1^u y_3^u}{\sqrt{v_1^2 + v_2^2}} \\
0 & \sum_j v_j^2 (y_2^u)^2 + (y_1^u)^2 v_s^2 & \frac{(v_1^2 - v_2^2) v_s y_1^u y_3^u}{\sqrt{v_1^2 + v_2^2}} \\
-2\frac{v_1 v_2 v_s y_1^u y_3^u}{\sqrt{v_1^2 + v_2^2}} & \frac{(v_1^2 - v_2^2) v_s y_1^u y_3^u}{\sqrt{v_1^2 + v_2^2}} & \sum_j v_j^2 (y_3^u)^2
\end{pmatrix}  , }
\end{gathered}
\end{equation}
\begin{equation}
\begin{gathered} 
\widetilde{\bf H}_d  = 
\begin{pmatrix}
(y_1^d)^2 v_s^2 & 0 & \frac{ { (v_2^2-v_1^2)} v_s y_1^d y^{d}_3 e^{-i\delta_3}}{\sqrt{v_1^2+v_2^2}}  \\
0 & (y_1^d)^2 v_s^2 & \frac{2v_1 v_2 v_s y_1^d y^{d}_3 e^{-i\delta_3}}{\sqrt{v_1^2+v_2^2}} \\
\frac{ {(v_2^2-v_1^2)} v_s y_1^d y^{d}_3 e^{i\delta_3}}{\sqrt{v_1^2+v_2^2}}  &  \frac{2v_1 v_2 v_s y_1^d y^{d}_3 e^{i\delta_3}}{\sqrt{v_1^2+v_2^2}} & \sum_j v_j^2 (y_3^d)^2 
\end{pmatrix}  +
\begin{pmatrix}
0 & 0 & 0 \\
0 & \sum_j v_j^2 (y_2^d)^2  &  \sqrt{\sum_j v_j^2} v_s y_2^d y_4^{d} e^{-i\delta_4} \\
0 &  \sqrt{\sum_j v_j^2} v_s y_2^d y_4^d e^{i\delta_4} & v_s^2 |y_4^d|^2
\end{pmatrix} .
\end{gathered}
\end{equation}
\end{widetext}
Realize how in this basis, the quark mass matrix with a positive sign, as in Eq.~\eqref{eq:Mu} compared to Eq.~\eqref{eq:Md}, in the initial (2,3) matrix element is the one without a sizeable mixing in the 2-3 sector, Eq.~\eqref{eq:Hu}. This explains why the original setup~\cite{Das:2019itj} cannot correctly fit the Jarlskog invariant. As if one uses it, the large size of the (3,3) matrix element (approximately given by $m_t^2$) compared to the (2,3) element provides a negligible contribution, whereas in our setup the (3,3) matrix element from the down quark sector is sufficiently close to the (2,3) element to end up giving the right size contributions to the mixing angle, ${ {\theta}_{13}^q } \sim {\cal O}(1)\beta^3$.

Let us first compute everything for the up quarks. Adding approximations to the picture can help us to determine the size of the off-diagonal elements and how strong their role is. Therefore, we obtain {
\begin{align} 
\widetilde{\bf H}_u &  \sim
\begin{pmatrix}
{\cal O} (m_u^2) & 0 & {\cal O} (m_u) m_t   \tan \beta \\
0 & {\cal O} (m_c^2) &  {\cal O} (m_u) m_t \\
{\cal O} (m_u) m_t   \tan \beta& {\cal O} (m_u) m_t  & m_t^2
\end{pmatrix}  \;,  
\end{align}   }
where it is possible to see that the contributions to the $1-3$ { and $2-3$  mixing sectors} have the size { $\tan \beta {\cal O} (m_u)/m_t \sim 10^{-7}$ and $ {\cal O} (m_u)/m_t \sim 10^{-5}$}, which { are rather negligible compared to those to come from the down quark sector.}

The down quark sector is a bit more complicated. There, we obtain
\begin{align} \notag
\frac{\widetilde{\bf H}_d }{ \beta^4 v^2} \sim &
\begin{pmatrix}
	\beta^8  (\widehat{y}_1^d)^2 & 0 & -\beta^4 \widehat{y}_1^d \widehat{y}_3^{d} \\
	0 & \beta^8   (\widehat{y}_1^d)^2 & 2 \beta^5 \widehat{y}_1^d \widehat{y}_3^{d} \\
	-\beta^4 \widehat{y}_1^{d} \widehat{y}_3^{d} & 2\beta^5 \widehat{y}_1^{d} \widehat{y}_3^{d} & (\widehat{y}^d_3)^2
\end{pmatrix} \\
 & +
\begin{pmatrix}
	0 & 0 & 0 \\
	0 & \beta^4   (\widehat{y}_2^d)^2 & \beta^2 \widehat{y}_2^d \widehat{y}_4^{d} \\
	0 & \beta^2 \widehat{y}_2^{d} \widehat{y}_4^{d} & (\widehat{y}^d_4)^2
\end{pmatrix} \;,
\end{align}
where we have substituted the order 1 Yukawa couplings which are defined as
\begin{equation}
\begin{gathered}
y_1^d = \beta^4 \widehat{y}_1^d \;,  \quad  y_2^d = \beta^4 \widehat{y}_2^d \;, \\
 y_3^d = \beta^2 \widehat{y}_3^d \;, \quad \text{and} \quad y_4^d = {\cal O}(1) \widehat{y}_4^d \;,
\end{gathered}
\end{equation}
and used
\begin{equation}
v_1 \simeq v\;, \quad v_2 \simeq \beta v\;, \quad \text{and} \quad v_s \simeq \beta^2 v \;.
\end{equation}

It is straightforward to estimate the order of magnitudes of the dominant contributions to the quark mixing angles:
\begin{align}
\theta_{23}^q \sim \beta^2 {\cal O}(1) \qquad
\text{and} \qquad 
\theta_{13}^q \sim \beta^4 {\cal O}(1) \;,
\end{align}
providing also the correct order of magnitude of the down quark masses
\begin{align}
m_d \sim {\cal O} (1) \beta^4 v  \;, \quad
m_s \sim {\cal O} (1)\beta^3 v  \;, \quad
m_b \sim {\cal O} (1)\beta^2 v \;.
\end{align}
Furthermore, note that both mixing angles are directly proportional to $v_s$ meaning that they must be proportional to $\beta$ as wanted.

\section{$\chi^2$-FIT IN THE QUARK SECTOR}
\label{app:chiQuarks}
\noindent   
In the following,  we search for the best-fit values within our set of nine  parameters
$\{y^u_1, y^u_2,y^u_3,y^d_1,y^d_2,y^d_3,y^d_4,\delta_3^d,\delta_4^d\}$ by virtue of a $\chi^2$ 
fit to the five quark masses $\{m_c,m_t,m_d,m_s,m_b \}$ and four mixing parameters
\begin{equation}
\begin{split}
\chi ^{2} = &\sum_{f=c,t,d,s,b}\frac{(m_{f}^{\text{th}}-m_{f}^{\text{exp}})^{2}}{\sigma_{f}^{2}}
+\frac{(|\mathbf{V}_{12}^{\text{th}}|-|\mathbf{V}_{12}^{\text{exp}}|)^{2}}{\sigma _{12}^{2}}
\\
& +\frac{(|\mathbf{V}_{23}^{\text{th}}|-|\mathbf{V}_{23}^{\text{exp}}|)^{2}}{\sigma _{23}^{2}}+\frac{(|\mathbf{V}_{13}^{\text{th}}|-|\mathbf{V}_{13}^{\text{exp}}|)^{2}}{\sigma _{13}^{2}}
\\
& + \frac{(J_{q}^{\text{th}}-J_{q}^{\text{exp}})^{2}}{\sigma _{J}^{2}}\, \;,
\end{split}
\end{equation}
where the value of the masses is taken at the $Z$ boson mass scale, $M_Z$, 
using the \texttt{RunDec} package~\cite{Herren:2017osy}
\begin{align}
\begin{split}
	m_c^{\text{exp}} (M_Z) & = 0.626 \pm 0.020 \text{ GeV} \;, \\
    m_t^{\text{exp}} (M_Z) & = 172.29 \pm 0.06 \text{ GeV} \;, \\
    m_d^{\text{exp}} (M_Z) & = 0.0027^{+0.0003}_{-0.0002} \text{ GeV} \;, \\
    m_s^{\text{exp}} (M_Z) & = 0.055^{+0.004}_{-0.002} \text{ GeV} \;, \\
    m_b^{\text{exp}} (M_Z) & = 2.86 \pm 0.02 \text{ GeV}  \;,
\end{split} 
\end{align}
and
\begin{align}
\begin{split}
    |\mathbf{V}_{12}^{\text{exp}}| & = 0.22452 \pm 0.00044 \;, \\
    |\mathbf{V}_{23}^{\text{exp}}| & = 0.04214 \pm 0.00076  \;, \\
    |\mathbf{V}_{13}^{\text{exp}}| & = 0.00365 \pm 0.00012 \;,\\
    J_{q}^{\text{exp}} & = (3.18\pm 0.15) \times 10^{-5}\;,
\end{split}
\end{align}
as shown in the most recent global fit from the PDG~\cite{Tanabashi:2018oca}.
Furthermore, we have considered: 
\begin{equation} { 
v_1 \simeq 174.0 \text{ GeV}, \quad v_2 \simeq 8.14 \text{ GeV}, \quad
\text{and}\quad v_s \simeq 2.63 \text{ GeV}. }
\end{equation}
The best-fit values are shown in Eq.~\eqref{eq:BFP}.

\section{LEPTONIC MASSES AND MIXING}
\label{app:chiLeptons}
\noindent   
The value of the charged lepton masses is taken at the $Z$ boson mass scale, $M_Z$, from Ref.~\cite{Antusch:2013jca}
\begin{align}
\begin{split}
	m_\tau^{\text{exp}} (M_Z) & = 1744.614156  \text{ MeV} \;, \\
    m_\mu^{\text{exp}} (M_Z) & = 102.627051 \text{ MeV} \;, \\
    m_e^{\text{exp}} (M_Z) & = 0.4861410527 \text{ MeV} \;, 
\end{split} 
\end{align}
the neutrino masses 
\begin{align}
\begin{split}
    \Delta m_{21}^2 & = 7.39^{+0.21}_{-0.20} \times 10^{-5} \text{ eV}^2 \;, \\
  \text{NO:  }  \Delta m_{31}^2 & = +2.528^{+0.033}_{-0.031} \times 10^{-3} \text{ eV}^2  \;\\
  \text{IO:  }  \Delta m_{32}^2 & = -2.510^{+0.032}_{-0.031} \times 10^{-3} \text{ eV}^2 \;,
\end{split}
\end{align}
and mixing parameters from~\cite{Esteban:2018azc}
\begin{align}
\begin{split}
    |\mathbf{U}_{12}^{\text{exp}}| & \in [0.518,0.585] \;, \\
    |\mathbf{U}_{13}^{\text{exp}}| & \in [0.143,0.156]  \;, \\
    |\mathbf{U}_{23}^{\text{exp}}| & \in [0.651,0.772] \;,\\
  \text{NO:  }  \delta_{CP}^{\text{exp}} & \in [144,357]^\circ \;, \\
  \text{IO:  }  \delta_{CP}^{\text{exp}} & \in [205,348]^\circ
\end{split}
\end{align}
where all the mixing parameters were given in their $3 \sigma$ range. The Dirac phase can be also translated to a value of the Jarlskog invariant, namely: $J_\ell = -0.0329$.

\bibliographystyle{apsrev4-1}\vspace{0cm}
\bibliography{3hdmtanbeta}

\end{document}